\documentclass[a4paper, twocolumn, superscriptaddress, longbibliography]{revtex4-2}
\usepackage[colorlinks,urlcolor=blue,linkcolor=blue]{hyperref}
\usepackage[utf8]{inputenc}
\usepackage{graphicx}
\usepackage{geometry}
\usepackage{amsmath}
\usepackage{amssymb}
\usepackage{amsfonts}
\usepackage{hyperref}
\usepackage{dsfont}
\usepackage[dvipsnames]{xcolor}
\usepackage{appendix}

\hypersetup{citecolor=cyan}
\newcommand{\vra}{\rangle \! \rangle}
\newcommand{\vla}{\langle \! \langle}
\newcommand{\hh}[1]{\hat{\hat{#1}}}
\newcommand{\kket}[1]{|#1\rangle\!\rangle}
\newcommand{\bbra}[1]{\langle\!\langle#1|}
\newcommand{\bbrakket}[2]{\langle\!\langle#1|#2\rangle\!\rangle}
\newcommand{\norm}[1]{|\!| #1 |\!|}

\begin{document}
	\author{Maxime Debertolis}
	\email{maxime.debertolis@ijs.si}
	\affiliation{Institute of Physics, University of Bonn, Nu\ss allee 12, 53115 Bonn}
	\title{Super natural orbital representation of many-body operators: structured non-Gaussianity and matrix product operator compression}

	\begin{abstract}
	We introduce super natural orbitals (SNOs) for many-body operators, defined as the eigenvectors of the one-body super-density matrix associated with an operator (OBDMO). These objects provide a natural measure of the complexity of operators in terms of non-Gaussianity. We first establish analytical properties of SNOs for time-evolution operators generated by non-interacting Hamiltonians and for Haar-random unitaries. We then perform numerical tensor network simulations to compute the SNOs for both the time-evolution operator and local operators in the Heisenberg picture in two many-body systems: the fermionic $t\text{-}V$ chain and a quantum impurity model. While the $t\text{-}V$ model exhibits no preferred super-orbital basis, the operators in the impurity model display exponentially decaying SNO occupations at all times, indicating that only a few SNOs contribute significantly to quantum correlations. For local Heisenberg-picture operators in the impurity model, we find that the complexity in the SNO basis saturates at long times. Finally, we show that rotating operators into the SNO basis with an appropriate ordering leads to substantial matrix product operator compression by exposing the factorized structure of a large number of SNOs.

	\end{abstract}

	\maketitle

	\section{Introduction}

	Strong correlations between particles, arising from their interactions, are at the heart of a wide array of exotic phenomena in condensed matter physics, ranging from
	high-temperature superconductivity~\cite{Keimer_Kivelson_Norman_Uchida_Zaanen_2015} to quantum spin liquids~\cite{Savary_2017} and many-body
	localization~\cite{Nandkishore_Huse_2015,Abanin_Altman_Bloch_Serbyn_2019}. Understanding the physics of such strongly correlated systems is notoriously difficult, as it
	typically requires handling exponentially large Hilbert spaces -- especially in non-perturbative regimes where analytic approaches fail. To overcome these challenges, a wide range
	of numerical techniques have been developed over the past decades, including exact diagonalization~\cite{Lanczos1952SolutionOS,Benner_fassbender_1997}, quantum Monte
	Carlo~\cite{Sandvik_Kurkijarvi_1991,Foulkes_Mitas_Needs_Rajagopal_2001}, and tensor network algorithms~\cite{Schollwock_2011,Banuls_2023}. Among these, tensor network
	methods -- particularly due to the efficient and practical matrix product states (MPS) and matrix product operators (MPO) representations -- have become a method of choice for
	tackling one-dimensional systems where entanglement remains sufficiently bounded.

	Operators play a central role in all these approaches. They encompass observables used to extract physical quantities, time-evolution operators that govern the dynamics of
	states or operators (in either the Schrödinger or Heisenberg picture), and the Hamiltonian itself, which is the cornerstone of simulations such as density matrix renormalization
	group (DMRG)~\cite{White_92,Ostlund_Rommer_1995,Schollwock_2011}. Moreover, operators can encode information about a quantum system that is inaccessible from the states
	alone -- for instance, through out-of-time-order correlators (OTOCs), which provide insight into the scrambling of quantum information across space and
	time~\cite{Larkin_1969,Maldacena_Shenker_Stanford_2016,Bohrdt_2017,Hemery_Pollmann_Luitz_2019,Xu_Swingle_2020,Gisti_Luitz_Debertolis_2025}. 

	In tensor network simulations, operators are typically encoded as MPOs. The efficiency of such representations depends critically on the internal correlations of the operator:
	higher correlations translate into larger bond dimensions and computational overhead. This is especially true for unitary operators, such as the time-evolution operator or
	time-evolved local operators, whose von Neumann entanglement entropy tends to grow rapidly -- following either a volume or area law -- thus making their simulation particularly
	challenging~\cite{Calabrese_2005,Fagotti_Calabrese_2008,Eisert_Cramer_Plenio_2010,Alba_2024,Murciano_Dubail_Calabrese_2024,Gisti_Luitz_Debertolis_2025}.  This growth is even more severe than in the MPS case, as the operator acts on a doubled Hilbert space, effectively squaring the system size and amplifying entanglement barriers.

	A crucial but often under-explored aspect of MPS and MPO simulations is the choice of the one-particle basis. The complexity of these tensor network representations is not intrinsic to the physical system alone, but also depends on how the degrees of freedom are organized. Specifically, it is desirable to identify an orbital basis in which the correlations and hence the entanglement between subsystems are minimized. Recent advances have demonstrated that, for specific problems such as quantum impurity models, the entanglement growth during a quench can be overcome by implementing a suitable time-dependent basis rotation~\cite{Nunez_Debertolis_Florens_2025}. In these systems, consisting of a strongly interacting impurity coupled to a large non-interacting fermionic bath, long-range correlations emerge naturally. 
Various strategies have been developed to reduce the complexity of quantum impurity problems through effective bath representations, including variational Gaussian states with symmetry-based decoupling transformations~\cite{Ashida_2018, PhysRevB.98.024103}, or optimization of the  bath degrees of freedom through the influence functionals as effective temporal MPS~\cite{Park_2024, Thoenniss_2025}. In contrast, it has been shown that expressing states in terms of natural orbitals -- eigenvectors of the one-body density matrix -- enables significant compression, due to the exponential decay of their occupation in both ground states~\cite{Zgid_Chan_2011,He_Lu_2014,Lu_Hoppner_Gunnarsson_Haverkort_2014,Bravyi_Gosset_2017,Boutin_Bauer_2021,Debertolis_Florens_Snyman_2021,Debertolis_Snyman_Florens_2022} and certain out-of-equilibrium settings~\cite{Nunez_Debertolis_Florens_2025}. Natural orbitals have also proven useful in characterizing many-body localized systems, where different basis choices reveal distinct one-particle properties~\cite{Bera_Schomerus_Meisner_Bardarson_2015,Bera_Martynec_Schomerus_Meisner_Bardarson_2017,Buijsman_Gritsev_Cheianov_2018,Hopjan_Meisner_2020}. 

	Building on this idea, we explore whether similar reductions in complexity can be achieved directly at the level of operators by introducing the concept of \textit{super natural orbitals} (SNOs) associated to operators. We define these as the eigenvectors   of the \textit{one-body density matrix of operators} (OBDMO), a natural generalization of the concept used for quantum states. Inspired by the structure of many-body states in particular physical contexts, we investigate the properties of unitary operators in spin chains and quantum impurity problems, expecting that the behavior of their  SNOs will vary significantly between these two settings. Our main result shows that, in quantum impurity models, the complexity of MPO representations could in principle often be exponentially reduced by working in the super natural orbital basis of the operator. This suggests that the compression seen in specific states (e.g., ground states) actually stems from fundamental properties of operator themselves -- properties that extend across the full many-body Hilbert space.

	Beyond its utility in quantifying and potentially reducing the complexity of classical simulations, the framework presented here also proves valuable for characterizing computational complexity in quantum simulations. A central topic of recent research has been the classification and quantification of quantum operations based on their resource requirements~\cite{Chitambar_Gour_2019}. According to the Gottesman-Knill theorem, Clifford circuits can be simulated efficiently on classical computers, as they require only polynomial resources~\cite{Gottesman_1998,Gottesmman_Aaronson_2004}. However, these circuits cannot generate all quantum states. To achieve universal quantum computation, non-Clifford gates must be introduced, which significantly increases the difficulty of classical simulations~\cite{Barenco_Bennett_etal_1995,Bravyi_Kitaev_2005}. The complexity introduced by such non-Clifford operations is captured by the concept of non-stabilizerness~\cite{Veitch_2014,Bravyi_Gosset_2016,Bravyi_Smith_Smolin_2016,Hamaguchi2024handbookquantifying}, also known as quantum magic, which has also be extended to fermionic systems~\cite{Collura_DeNardis_Alba_Lami_2025,Sierant_Stornati_Turkeshi_2025}. Quantifying this resource demands efficient and practical methods for classical computation~\cite{Leone_Oliviero_Hamma_2022,Haug_Piroli_2023,Lami_Collura_2023,Leone_Bittel_2024,Tarabunga_Tirrito_Banuls_Dalmonte_2024}. Similarly, fermionic Gaussian states -- eigenstates of non-interacting Hamiltonians or those generated by matchgate circuits -- are also amenable to efficient classical simulation~\cite{Terhal_DiVincenzo_2002,Bravy_2004_Lagrangian}. In contrast, the inclusion of interactions introduces non-Gaussian features, thereby increasing complexity as the underlying computational Fock space is exponentially large: this other kind of complexity is captured by the so-called measure of non-Gaussianity~\cite{ReardonSmith_Oszmaniec_Korzekwa_2024,Dias_Koenig_2024,Lyu_Bu_2024,Coffman_Smith_Gao_2025,Sierant_Stornati_Turkeshi_2025}. Both non-stabilizerness and non-Gaussianity become particularly meaningful in the context of operators, which represent quantum circuits or time-evolution operators associated to a given Hamiltonian~\cite{Bravy_2004_Lagrangian,Dowling_Kos_Turkeshi_2024}, and have been so far less investigated due to the lack of a framework for operators. In this work, we demonstrate that non-Gaussianity can be extended to many-body operators and efficiently computed using tensor network methods. We also explore the properties of this measure in the specific systems studied here.

	The paper is organized as follows: In Section~\ref{sec:NO_st}, we review the properties of the one-body density matrix (OBDM) for quantum states, focusing on the features we aim to generalize to operators. In Section~\ref{sec:definition}, we introduce the notion of  SNOs for operators and define orbital rotations in terms of super-orbitals in a doubled Hilbert space. We also analyze the properties of such rotations when the operator is unitary, and illustrate the formalism with a simple case of a factorized operator. We then discuss the measures of non-Gaussianity for operators, including mixed or pure state density matrices and unitary operators. 
	In Section~\ref{sec:special_cases}, we consider two limiting cases: non-interacting systems, where the natural orbital basis trivializes the MPO structure, and Haar-random unitaries, where no basis is preferred.  The main numerical results are presented in Section~\ref{sec:results}, where we investigate the time-evolution operator and time-evolved local observables in the $t$–$V$ model and in a quantum impurity model. We show that in the latter case, the MPO complexity can be significantly reduced using the SNO basis. Finally, in Section~\ref{sec:discussion}, we summarize our findings and outline possible future directions stemming from the concepts introduced in this work.

	\section{One-body density matrix and natural orbitals for states}
	\label{sec:NO_st}
	We start this article by reviewing the definition of the  one-body density matrix (OBDM) defined for an arbitrary state $|\psi\rangle$, and some of its properties that we extend for
	operators later on. For simplicity, we consider states with a well defined number of particles, but the definitions can be extended beyond this constraint. Given a fermionic system in which the particle can occupy $L$ orbitals , equipped with the
	fermionic creation and annihilation operators $\hat{c}^{\dagger}_{l}$ and $\hat{c}^{}_{l}$ respectively creating and removing a fermion in the orbital $l$, a state
	$|\psi\rangle$ can be written in a complete basis of the fermionic Fock space:
	\begin{equation}
		\label{eq:many_body_state}
		|\psi\rangle = \sum_{i=1}^{2^{L}} a^{}_{i}|i\rangle = \sum_{i=1}^{2^L} a^{}_{i}  \left[\prod_{l=1}^{L}\left(\hat{c}^{\dagger}_{l}\right) ^{n^{i}_l}\right]|0\rangle,
	\end{equation}
	where $a^{}_{i}$ are complex coefficients which for a normalized state are constrained as $\sum_i |a_i^{}|^{2} = 1$ and $|i\rangle$ are  Slater determinants and product states in the occupation basis forming a complete basis $\sum_i |i\rangle \langle i| = \mathds{1}$. $n^{i}_l=\langle i|\hat{n}^{}_{l}|i \rangle$ is the occupation of the orbital $l$ in the state $|i\rangle$ and $|0\rangle$ is the vacuum state.  The  OBDM, which we label here $Q$, is a $(L\times L)$ Hermitian matrix whose elements are defined  as :
	\begin{equation}
		Q^{}_{ij} = \langle \psi | \hat{c}^{\dagger}_{i}\hat{c}^{}_{j} | \psi \rangle.
	\end{equation}
	$Q$ being Hermitian, there exists a unitary  matrix $V$ such that $Q=VDV^{\dagger}$, such that $V$ corresponds to a rotation of the one-particle basis of orbitals $\hat{c}$ in
	the so-called natural orbitals basis $\hat{q}$. This basis rotation can be written as a linear combination of the  original set of orbitals  using the elements of $V$:
	\begin{equation}
		\hat{q}_{\alpha} = \sum_{i=1}^{L}V^{*}_{i\alpha} \hat{c}^{}_{i}, \quad \quad \hat{c}_{i} = \sum_{\alpha=1}^{L}V^{}_{i\alpha} \hat{q}^{}_{\alpha}.
	\end{equation}
	The  eigenvalues of $Q$ are the occupations of the natural orbitals $\hat{q}$ in the state $|\psi\rangle$, namely:
	\begin{equation}
		D_{\alpha,\beta} = \langle \psi|\hat{q}^{\dagger}_{\alpha}\hat{q}^{}_{\beta}|\psi\rangle \delta_{\alpha,\beta}^{} = n_{\alpha} \delta_{\alpha,\beta}^{}.
	\end{equation}
	 The natural orbitals have the property to maximize the sum of occupations of the $r$ most occupied orbitals  $\sum_{\alpha=1}^r n_\alpha$. This object is widely used in the quantum chemistry community to organize molecular orbitals according to their degree of
	correlations~\cite{Lowdin_1955,Davidson_1972,Roos_Taylor_Sigbahn_1980}. Therefore, if the state $|\psi\rangle$ can be represented as a  single Slater determinant in a given one-particle basis, it will be revealed by the natural orbital basis, whose occupations will be strictly $n_{\alpha}=1$ (occupied orbital) or $0$ (empty orbital). Correlations lead to occupations that are different from zero or one.

	\section{Super natural orbitals for operators}
	\label{sec:definition}

	The  OBDM defined above serves as a tool to assess the complexity of a given state by measuring its proximity to a  Slater determinant. We now extend this line of inquiry
	to operators: can we construct an analogous object that captures the complexity of an operator -- specifically, one that indicates whether the operator can be expressed  in a factorized form in a given basis? This question is particularly relevant for the numerical simulation of quantum systems in
	the Heisenberg picture, as discussed in the introduction.  The time evolution of operators can be performed using matrix product operators (MPOs), whose complexity
	depends on the chosen basis. A notable example is that of  non-interacting systems,  in which the time-evolution operator can be expressed as an MPO with bond dimension  $\chi_i=1$ in the orbital
	basis  diagonalizing the Hamiltonian matrix,  while the MPO representation breaks down for a simulation in real space. 

	In order to follow the same derivation as before,  we \textit{vectorize} the operators such that their mathematical formulation is similar to the one for states. The vectorization of an operator $\hat{O}$ is  defined through the following operator-state correspondence:
	\begin{equation}
		\label{eq:def_op}
		\begin{split}
			\hat{O} &= \sum_{i,j=1}^{2^L} O^{}_{ij}|i\rangle\langle j|, \\
			|\hat{O}\vra &:= \sum_{i,j=1}^{2^L} O^{}_{ij}|i\rangle|j\rangle.
		\end{split}
	\end{equation}
	Using the Hilbert-Schmidt norm of the operator $\norm{\hat{O}}^{2} = \bbrakket{\hat{O}}{\hat{O}} = \mathrm{Tr}(\hat{O}^{\dagger}\hat{O})$, the vectorized operator $\kket{\hat{O}}/\norm{\hat{O}}$ becomes a normalized pure state in the doubled Hilbert space. Following the form of a state as defined in Eq.~\eqref{eq:many_body_state}, we can write an arbitrary operator $\hat{O}$ using creation and annihilation operators:

	\begin{equation}
		\label{eq:vec_op}
		\begin{split}
			\hat{O}& = \sum_{i,j=1}^{2^L} O^{}_{ij} \left[\prod_{l=1}^{L} \left(\hat{c}^{\dagger}_{l}\right) ^{n^{i}_l}\right]|0 \rangle
			\langle 0| \left[\prod_{l=L}^{1} \left(\hat{c}^{}_{l}\right)^{n^{j}_l} \right], \\
			|\hat{O}\rangle\!\rangle &= \sum_{i,j=1}^{2^L} O^{}_{ij} \left[\prod_{l=1}^{L} \left(\hat{c}^{\dagger}_{l}\right) ^{n^{i}_l}\right]\otimes\left[\prod_{l=1}^{L}
			\left(\hat{\tilde{c}}^{}_{l}\right)^{n^{j}_l} \right]^{T}\!\!\!|0 \vra,
		\end{split}
	\end{equation} 
	where the transpose is taken on the operators acting on the dual space labelled with a tilda, since they act from the left  in the first line of Eq.~\eqref{eq:vec_op}. The vacuum state
	is the tensor product of the vacua of the direct and dual spaces, $|0\vra = |0\rangle \otimes |0\rangle$, or equivalently the vacuum of a \emph{super-space} corresponding to a doubled
	Hilbert space $\mathcal{H}\otimes\mathcal{H}$. We can simplify the form of Eq.~\eqref{eq:vec_op}, by introducing the following \emph{super-operators} acting on the doubled Hilbert
	space, which we label with a double hat:
	\begin{equation}
		\label{eq:superop}
		\hat{\hat{d}}^{\dagger}_{m} = 
		\begin{cases}
			\hat{c}^{\dagger}_{m} \otimes \mathds{1} \hspace{1.45cm} \text{if } \quad m\leq L \\
			\mathds{1} \otimes \hat{\tilde{c}}^{}_{m-L} \hspace{1.cm} \text{otherwise}
		\end{cases},
	\end{equation}
	where $m\in[1,2L]$ and $\hat{\tilde{c}}$ means that the matrix representation of the operator is transposed. Super-operators within the same set (direct or dual) obey canonical anticommutation relations, while operators from different sets commute, reflecting the commutativity of left and right multiplication. A genuine fermionic algebra on $2L$ modes can be obtained by dressing the dual operators with the fermion parity as Klein factors (similarly to the third quantization of Ref.~\cite{prosen2008}), but it is not required in the present work. All objects defined in the following act separately on direct and dual sectors. Using these super-operators, the vectorized operator from Eq.~\eqref{eq:vec_op}  becomes:
	\begin{equation}
		\label{eq:state_like_operator}
		|\hat{O}\rangle\!\rangle = \sum_{k=1}^{4^{L}} O_{k}\left[\prod^{2L}_{m=1} \left(\hat{\hat{d}}_{m}^{\dagger}\right)^{n^{k}_{m}} \right]|0\rangle\!\rangle,
	\end{equation}
	which is now very similar to the state written in Eq.~\eqref{eq:many_body_state}. The former indices $i$ and $j$ from Eq.~\eqref{eq:vec_op} can be recovered from  the reshape of the
	matrix $O_{ij}$ of shape $(2^{L}\times 2^{L})$ into the vector $O_{k}$ of dimension $4^{L}$.  The vectorization can be represented graphically as follows: 
	\begin{figure}[h!]
		\centering
		\label{fig:vectorized_op}
		\includegraphics[width=1.\columnwidth]{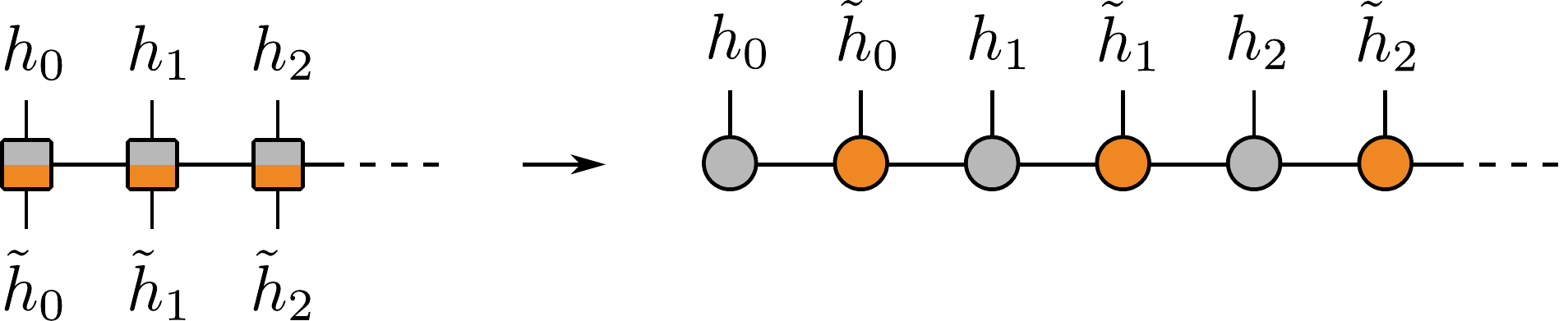}
	\end{figure} \\
	where $h_i$ (resp. $\tilde{h}_i$) refers to the local \emph{direct} (resp. \emph{dual}) Hilbert space at site $i$, associated with the gray (resp.  orange) color. 
  
	The introduction of the super-operators $\hat{\hat{d}}_m$ allows us to extend the definition of the  OBDM to operators in terms of their vectorized form and the
	$2L$ super-operators of Eq.~\eqref{eq:superop}. We call $R[\hat{O}]$ the $(2L\times2L)$ dimensional Hermitian  \emph{one-body density matrix of operators} (OBDMO) associated to  the operator $\hat{O}$, whose elements read:
	\begin{equation}
		\begin{split}
			R[\hat{O}]^{}_{m,n} &= \frac{ \vla \hat{O} | \hat{\hat{d}}^{\dagger}_{m} \hat{\hat{d}}_{n} | \hat{O} \vra }{\norm{\hat{O}}^{2}} \\
			&= \frac{1}{\norm{\hat{O}}^{2}} \vla \hat{O} | 
			\begin{pmatrix}
				\hat{c}^{\dagger}_{m}\hat{c}^{}_{n} \otimes \mathds{1} & \hat{c}^{\dagger}_{m}\otimes\hat{\tilde{c}}^{}_{n} \\
				\hat{c}^{}_{n}\otimes \hat{\tilde{c}}^{\dagger}_{m} & \mathds{1} \otimes \hat{\tilde{c}}^{}_{n}\hat{\tilde{c}}^{\dagger}_{m} 
			\end{pmatrix}
			|\hat{O} \vra,
		\end{split}
		\label{eq:operator_obrdm}
	\end{equation}
	which naturally takes the form of a matrix with four blocks, the off-diagonal ones mixing the direct and dual spaces of the operator. This definition does not account for anomalous terms such as $\hat{c}^{}_{m}\hat{c}^{}_{n}\otimes \mathds{1}$ or $\hat{c}^{\dagger}_{m}\hat{c}^{\dagger}_{n}\otimes \mathds{1}$, which are required for Hamiltonians that do not conserve particle number, as relevant to superconductivity. The extension to include such terms is straightforward and yields a $(4L\times 4L)$ matrix with a similar block structure, and can be defined by using third quantization super-majorana operators~\cite{prosen2008} instead of the algebra defined in Eq~\eqref{eq:superop}. For simplicity, we restrict the present discussion to number-conserving Hamiltonians and defer the general case to future work. It can also be generalized to bosonic degrees of freedom, by similarly extending the definition of the bosonic version of the OBDM for states~\cite{PhysRev.104.576, PhysRevLett.124.180603,Garc_a_March_2016} to operators. \\

	As for the  OBDM $Q$, the OBDMO $R$  is
	diagonalized by a unitary matrix $W$ such that $R = WDW^{\dagger}$,  corresponding to a rotation of the super-orbitals:
	\begin{equation}
		\label{eq:rot_op}
		\hat{\hat{q}}_{\alpha} = \sum_{m=1}^{2L}W^{*}_{m\alpha} \hat{\hat{d}}^{}_{m}, \quad \quad \hat{\hat{d}}_{m} = \sum_{\alpha=1}^{2L}W^{}_{m\alpha} \hat{\hat{q}}^{}_{\alpha}.
	\end{equation}
	We dub the super-orbitals $\hat{\hat{q}}$ the \emph{super natural orbitals} (SNOs).
	The specificity of the rotation $W$ compared to the previous rotation $V$ comes from the off-diagonal blocks of $R$, and is not simply a rotation of the system's orbitals
	$\hat{c}$ but also mixes the dual and direct spaces of the operator. The occupation $n_\alpha = \vla \hat{O}| \hat{\hat{q}}_\alpha^\dagger\hat{\hat{q}}_\alpha^{} |\hat{O}\vra/\norm{\hat{O}}^{2}$  of the SNOs are the eigenvalues of the OBDMO $R$. Since $\kket{O}/\norm{O}$ is a normalized state, $0\leq\lambda_\alpha \leq 1$ for arbitrary operators. In the following, we refer to these eigenvalues as the correlation spectrum of the operator. Using the graphical notation for tensors, the elements of  $R$  are computed as depicted in the following
	sketch: 
	\begin{figure}[h!]
		\label{fig:element_Rmn}
		\centering
		\includegraphics[width=1.\columnwidth]{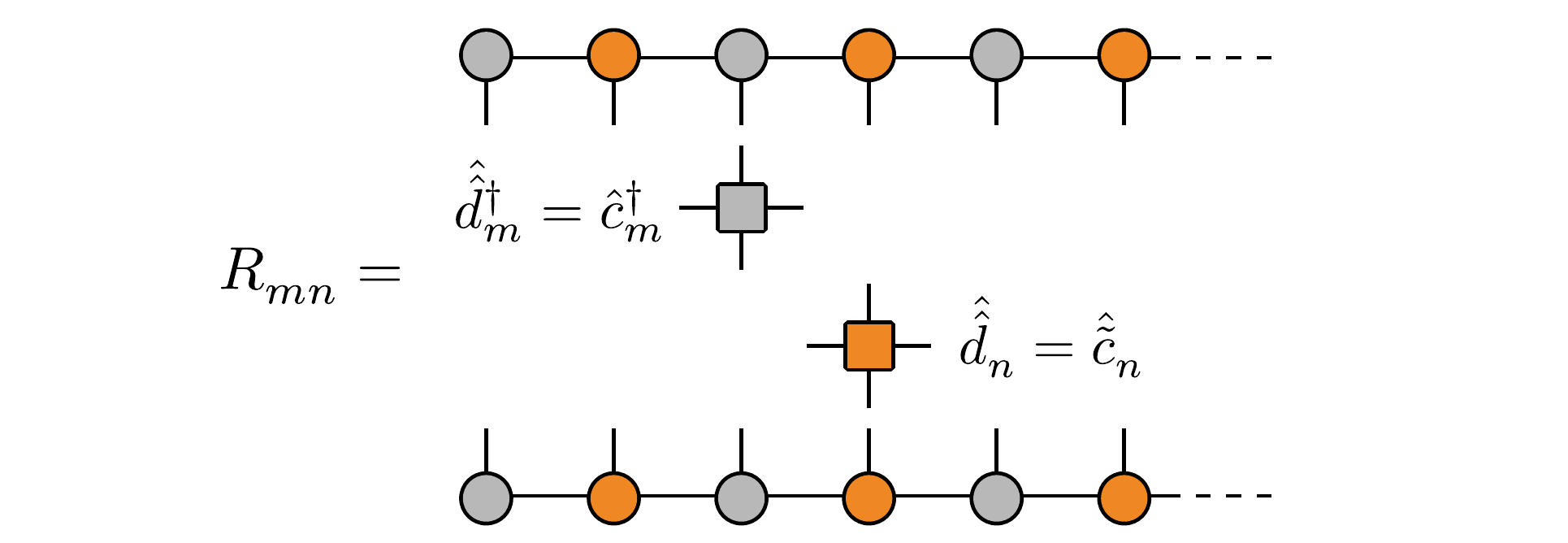}
	\end{figure} \\
	The  OBDMO as defined in Eq.~\eqref{eq:operator_obrdm} has been already introduced in ref.~\cite{Murciano_Dubail_Calabrese_2024,Rath_Vitale_Murciano_Votto_Dubail_Kueng_Calabrese_Vermersch_2023,Siva_Zou_Soejima_Mong_Zaletel_2022} as a conformal field theory tool
	to study bipartite entanglement properties, but no attention has been put into these objects  for measuring the complexity of operators. As  for the case of states, the
	 SNOs of  an operator can reveal if there exists (or not) a basis in which the operator  is in a factorized form, for which Eq.~\eqref{eq:state_like_operator} has only a single non-zero element. However, due to the mixing of the direct and dual spaces, the operator might not be an MPO anymore but an MPS in the doubled Hilbert space, unless the  SNO preserve the tensor product structure between the direct and the dual spaces. In practice, the  OBDMO does not require the vectorized form of the operator,  using:
	\begin{equation}
		\label{eq:trace_definition}
		\begin{split}
			\vla \hat{O} |\hat{c}^{\dagger}_{m}\hat{c}^{}_{n} \otimes \mathds{1} | \hat{O} \vra &= \mathrm{Tr}\left( \hat{O}^{\dagger}\hat{c}^{\dagger}_{m}\hat{c}^{}_{n}\hat{O}\right),  \\
			\vla \hat{O} |\hat{c}^{\dagger}_{m}\otimes\hat{\tilde{c}}^{}_{n} | \hat{O} \vra &= \mathrm{Tr}\left( \hat{O}^{\dagger}\hat{c}^{\dagger}_{m}\hat{O}\hat{c}^{}_{n}\right), \\
			\vla \hat{O} |\mathds{1} \otimes \hat{\tilde{c}}^{}_{m}\hat{\tilde{c}}^{\dagger}_{n} | \hat{O} \vra &= \mathrm{Tr}\left(
			\hat{O}^{\dagger}\hat{O}\hat{c}^{}_{m}\hat{c}^{\dagger}_{n}\right),  \\ 
		\end{split}
	\end{equation}
	such that elements of $R$ can be efficiently computed  as traces of products of MPOs, a common operation in tensor network simulations.  

	In the following, we develop with more details the cases of unitary operators and density matrices for which the OBDMO acquire specific structures, and introduce the correlation entropy as a measure of the operator complexity in terms of non-Gaussianity.

	\subsubsection*{Unitary operators}

	Unitary operators have the property that $\hat{O}^{\dagger} \hat{O} = \mathds{1}$ such that $\norm{\hat{O}}^{2} = \mathrm{Tr}(\mathds{1})$, leading to  simplifications in the OBDMO.  Starting from the definition in Eqs.~\eqref{eq:trace_definition} and by using the cyclicity of the trace, the diagonal blocks become:

	\begin{equation}
		\begin{split}
			\vla \hat{O} | \hat{c}^{\dagger}_{m}\hat{c}^{}_{n} \otimes \mathds{1} |\hat{O} \vra &=  \frac{\mathrm{Tr}(\mathds{1})}{2}\delta_{mn}, \\
			\vla \hat{O} | \mathds{1} \otimes \hat{\tilde{c}}^{}_{m}\hat{\tilde{c}}^{\dagger}_{n}|\hat{O} \vra &= \frac{\mathrm{Tr}(\mathds{1})}{2}\delta_{mn}, \\ 
		\end{split}
	\end{equation}
	where the trace of the identity  is compensated by the norm of the operator.  Thus, only off-diagonal blocks of the  OBDMO are  computed for unitary operators:
	\begin{equation}
		\label{eq:R_unitary}
		R = 
		\begin{pmatrix}
			\frac{1}{2}\mathds{1}_L^{} & \Delta \\
			\Delta^{\dagger} & \frac{1}{2}\mathds{1}_L^{} \\
		\end{pmatrix}
		= 
		\frac{\mathds{1}_{2L}^{}}{2} + 
		\begin{pmatrix}
			0 & \Delta \\
			\Delta^{\dagger} & 0 \\
		\end{pmatrix},
	\end{equation}
	where $\Delta_{mn} = \mathrm{Tr}(\hat{O}^{\dagger}\hat{c}^{\dagger}_{m}\hat{O}\hat{c}^{}_{n})/\mathrm{Tr}(\mathds{1})$ is the only non-trivial $(L\times L)$ block of $R$. This simplified structure allows to express the eigenvectors of $R$ using the singular value decomposition of $\Delta$. Let $\mu_\alpha^{}$ be the $\alpha$-th singular values of $\Delta$, $u_\alpha^{}$ the corresponding left-singular vector and $v_\alpha^{}$ the right singular vector. The eigenvalues of $R$ come by pairs as $\lambda_{\alpha}^{\pm} = \frac{1}{2}\pm|\mu_{\alpha}|$ and correspond to the occupations of the SNO. Positivity of $R$ follows from its Gram-matrix structure, $w^\dagger R w = \norm{\sum_m w_m \hat{\hat{d}}_m \kket{\hat{O}} }^{2}/\norm{\hat{O}}^{2} \geq 0$. The upper bound follows from the positivity of the complementary Gram matrix $G_{mn}=\bbra{\hat{O}} \hat{\hat{d}}_m^{} \hat{\hat{d}}^{\dagger}_n \kket{\hat{O}} / \norm{\hat{O}}^{2}$, which the commutation relations of the super-operators reduce to $G = \mathds{1} - \Sigma R^{T}\Sigma$, with $\Sigma = \text{diag}(\mathds{1}_L, - \mathds{1}_L)$ (coming from the algebra of Eq.~\eqref{eq:superop}): since $\Sigma$ is unitary and $R$ Hermitian, $\Sigma R^T \Sigma$ shares the spectrum of $R$, and hence  $0 \leq \lambda_\alpha^{\pm} \leq 1$. The eigenvectors of $R$ also come by pairs and are expressed through the left- and right-singular vectors of $\Delta$ as $W_{\alpha,\pm} = \frac{1}{\sqrt{2}}(u_\alpha^{}, \pm v_\alpha^{})$. The pairs of SNOs are thus defined as:
	\begin{equation}
		\label{eq:nat_orb_doubled_dec}
		\hh{q}^{\dagger}_{\alpha,\pm} = \sum_{i=1}^{L} \frac{1}{\sqrt{2}} \left(u_{i\alpha}^{*} \hat{c}^{\dagger}_{i} \pm v_{i\alpha}^{*}\hat{\tilde{c}}^{}_{i} \right),
	\end{equation}
	We notice that the local direct and dual spaces are mixed in a simple way in each SNO, namely $\hat{c}^{\dagger}_{i} \pm \hat{\tilde{c}}^{}_{i}$.  Hence, we perform a linear combination of SNOs such that direct and dual spaces do not mix:
	\begin{equation}
		\label{eq:nat_orb_dec}
		\hat{q}^{\dagger}_{\alpha,\mathrm{direct}} = \sum_{i=1}^{L} u_{i\alpha}^{*} \hat{c}^{\dagger}_{i} \quad ; \quad 
		\hat{q}^{\dagger}_{\alpha,\mathrm{dual}}   = \sum_{i=1}^{L} v_{i\alpha}^{*} \hat{\tilde{c}}^{}_{i} \,,
	\end{equation}
	where $\hat{q}_{\alpha, \text{direct}}^{} = (\hat{q}^{}_{\alpha,+} + \hat{q}^{}_{\alpha,-})/\sqrt{2}$ and $\hat{q}_{\alpha, \text{dual}}^{} = (\hat{q}^{}_{\alpha,+} - \hat{q}^{}_{\alpha,-})/\sqrt{2} $.
	The previous equation shows that the  SNOs of a unitary operator are  obtained by a left and a right orbital rotations of the operator, $\hat{U}^{}_{\text{L}} \hat{O}\hat{U}^{\dagger}_{\text{R}}$, justifying the algebra defined in Eq.~\eqref{eq:superop}. When the matrix $\Delta$ is normal ($\Delta\Delta^{\dagger} = \Delta^{\dagger}\Delta$) -- which is the case for Hermitian operators $\hat{O} = \hat{O}^{\dagger}$ -- the left and right rotations coincide $\hat{U}_\text{L}^{} = \hat{U}_\text{R}^{}$.   The fact that SNOs of unitary operators do not mix direct and dual spaces imply that the MPO structure is preserved after the left and right rotations. The following sketch illustrates the two types of rotation for a	generic operator (left) and for unitary operators (right):
	\begin{figure}[h!]
		\label{fig:sketch_rotation}
		\centering
		\includegraphics[width=1.\columnwidth]{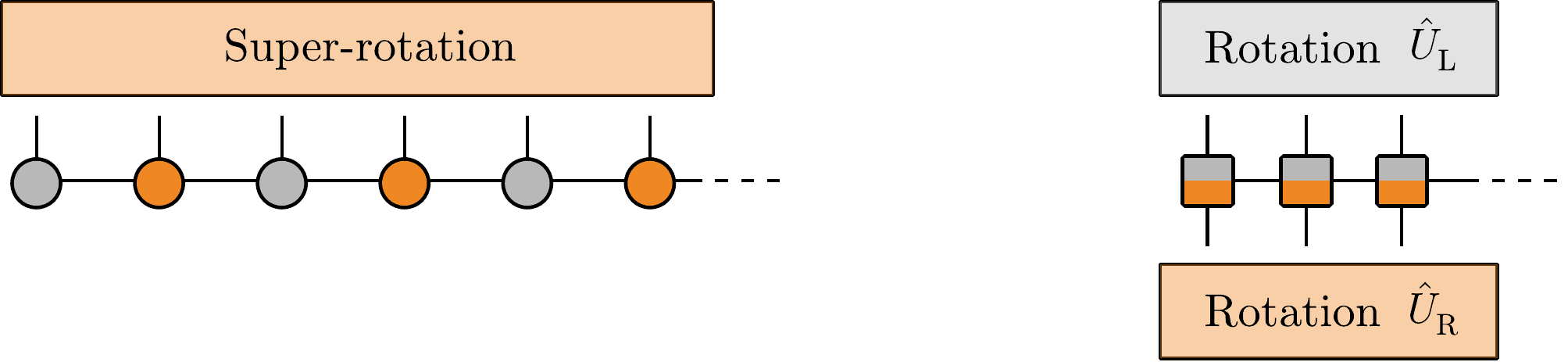}
	\end{figure} 

	\subsubsection*{Density matrices}

	We focus now on density matrices, especially for pure states  represented by rank one matrices. In this case,  the OBDMO is equivalent to the OBDM for the corresponding state. The density matrix of a system in a pure state $|\psi\rangle$ can be written and vectorized:
	
	\begin{equation}
		\hat{\rho} = |\psi\rangle\langle \psi| \rightarrow |\hat{\rho}\vra = |\psi\rangle\otimes|{\bar{\psi}}\rangle.
	\end{equation}
	with the normalization $\norm{\hat{\rho}}^{2}=1$. The corresponding OBDMO written by blocks involving the original fermion operators reads: 
	\begin{equation}
		R = 
		\begin{pmatrix}
			\langle \psi | \hat{c}^{\dagger}_{m}\hat{c}^{}_{n}|\psi\rangle 
			& \langle \psi|\hat{c}^{\dagger}_{m}|\psi\rangle \langle {\psi}|\hat{\tilde{c}}_{n}|{\psi}\rangle \\ 
			 \langle\psi|\hat{c}^{}_{n}|\psi\rangle \langle{\psi}|\hat{\tilde{c}}^{\dagger}_{m}|{\psi}\rangle
			& 1 - \langle {\psi}|\hat{\tilde{c}}^{\dagger}_{m}\hat{\tilde{c}}^{}_{n} |{\psi}\rangle
		\end{pmatrix}.
	\end{equation}
	The terms not preserving the particle number like $\langle\psi|\hat{c}^{\dagger}_{n}|\psi\rangle$ are simply $0$, such that:
	\begin{equation}
		R[\rho] = 
		\begin{pmatrix}
			Q[{\psi}] & 0 \\
			0 & 1-{Q}[{\psi}].
		\end{pmatrix}
		\label{eq:dens_mat}
	\end{equation}

	We emphasize that the block-diagonal form of Eq.~\eqref{eq:dens_mat} holds for pure states only. For mixed states $\hat{\rho} = \sum_i a_i|\psi_i \rangle\langle \psi_i|$, no such factorization occurs, and the off-diagonal matrix $\Delta$ receives contributions from pairs of eigenstates in adjacent charge sectors. The norm of a mixed density matrix coincides with its purity, $\norm{\hat{\rho}}^{2}=\mathrm{Tr}(\hat{\rho}^{2})$. Since the two matrix elements $\sum^{}_{i\neq j} a_i a_j \langle\psi_i|\hat{c}^{\dagger}_m|\psi_j\rangle\langle\psi_j|\hat{c}^{}_{n}|\psi_i\rangle$ change the particle number in opposite directions, these terms are compatible with $[\hat{\rho}, \hat{N}]=0$ and are generically nonzero. The direct and dual spaces do mix for mixed states, and the OBDMO then contains strictly more information than the OBDM using a single copy of $\hat{\rho}$. These off-diagonal blocks are thus essential: for a Gaussian mixed state whose NO occupations lie anywhere in $[0,1]$ reflecting purity rather than non-Gaussianity, which is the known conflation of covariance-based quantities~\cite{Sierant_Stornati_Turkeshi_2025, haug2026}, they restore SNO occupations exactly equal to $0$ and $1$, such that the correlation entropy defined below correctly vanishes, which can independently be related to the mixed-state non-Gaussianity defined in Ref.~\cite{tarabunga2026cov}. The simplest illustration is a single mode with $\hat{\rho}=(1-p)|0\rangle\langle 0| + p|1\rangle\langle 1|$, for which the off-diagonal element equals $p(1-p)$ and the two eigenvalues of $R$ are exactly $0$ and $1$ for any $p$, while the OBDM occupation gives $p$.
\\

 We can show the same results by using the definition of the elements of $R$ through the trace of operators as defined in Eq.~\eqref{eq:trace_definition} equivalently for density matrices. A pure state holds the following properties: $\hat{\rho}^{\dagger} = \hat{\rho}$ and $\hat{\rho}^{\dagger}\hat{\rho} = \hat{\rho}$. In this case, only the diagonal blocks of the $R$ matrix are non-zero contrary to unitary operators and they read:
	\begin{equation}
		\mathrm{Tr}\left( \hat{\rho} \hat{c}^{\dagger}_{m}\hat{c}^{}_{n}\right) = \langle \psi|\hat{c}^{\dagger}_{m}\hat{c}^{}_{n}|\psi\rangle 
	\end{equation}

	\subsubsection*{Correlation entropy}

	 From the occupations of SNOs, we can define a single measure of the complexity of an operator . This measure is commonly referred to as the correlation entropy for states, and reads:
	\begin{equation}
		\label{eq:corr_ent}
		S_{\mathrm{corr}}^{} = - \sum_{\alpha=1}^{2L}n^{}_{\alpha} \log\left( n^{}_{\alpha} \right).
	\end{equation}
	When the operator can be factorized  in a basis, the occupations of the  SNO are either $0$ or $1$, and the correlation entropy  is $S_{\mathrm{corr}}=0$. In the opposite case, the furthest from a  factorized form an operator is, the more  SNOs have their occupation $n_{\alpha}$ close to $\frac{1}{2}$, and the correlation entropy will be maximal when  the occupation of every SNO reaches this value, leading to the extensive maximal correlation entropy $S_{\mathrm{max}} = L\log(2)$. We point out here that the spectrum of  the OBDMO does not depend on the basis in which the operator is defined, such that operators  related through  a super-orbital rotation (in the sense defined above)  have the same  correlation entropy. \

	\subsubsection*{Operator non-Gaussianity}
	\textit{Non-Gaussianity}: The correlation entropy defined above serves as a measure for the operator complexity in terms of  non-Gaussianity. It is a faithful resource, meaning that $S_{\mathrm{corr}} = 0$ if and only if $\hat{O}$ factorizes into a product of operators each acting on a single SNO. This case contains the Gaussian forms -- $\hat{O} = \exp(M_{mn}^{}\hat{c}^{\dagger}_{m}\hat{c}^{}_{n})$  where repeated indices are implicitly summed  for any matrix $M$ -- together with their singular limits: projectors like $\hat{n}_i$ (up to normalization), or pure states, which are not exponentials of one-body operators themselves. As discussed earlier, the extension to terms breaking the particle number conservation $\hat{c}^{\dagger}_m \hat{c}^{\dagger}_n$ and $\hat{c}^{}_{m} \hat{c}^{}_{n}$ is also possible but is not discussed here for simplicity. Such an operator corresponds to a basis rotation, i.e., $\hat{O}^{\dagger} \hat{c}_{m} \hat{O} = \sum^{L}_{n=1}W_{mn}\hat{q}^{}_{n}$, with $W = \exp(M)$ defining an invertible linear transformation of the modes.This measure is also invariant under Gaussian operations: for any Gaussian operator  $\hat{U}_{\text{G}}^{}$,  the correlation entropy satisfies  $S_{\mathrm{corr}}(\hat{U}_{\text{G}}^{\dagger} \hat{O}\hat{U}_{\text{G}}^{}) = S_{\mathrm{corr}}(\hat{O})$, making it a basis-independent quantity.

	The faithfulness can be seen with the following argument: the vectorized operator $\kket{\hat{O}} = (\hat{O} \otimes \mathds{1}) \kket{\mathds{1}}$ is a pure state for any $\hat{O}$ by definition and the vectorized identity $\kket{\mathds{1}}$ is a Gaussian state of the super-modes. If $\hat{O} = \exp( M^{}_{mn} c^{\dagger}_{m}c^{}_{n}$, then $\hat{O}\otimes \mathds{1}$ is the exponential of a one-body super-operator, which maps Gaussian states to Gaussian states: any linear combination of super-operators annihilating $\kket{\mathds{1}}$ is conjugated into another linear combination annihilating $\kket{\hat{O}}$. The one-body density matrix of a pure Gaussian state is a projector, hence the occupations are $0$ or $1$ and $S^{}_{\mathrm{corr}}=0$. Conversely, a projector OBDMO of a pure state forces $\kket{\hat{O}}$ to be a (generatlized) Slater determinant of super-orbitals, which devectorizes into the factorized form.

is Gaussian, conjugation by $\exp(\text{quadratic})$ maps linear operators to linear operators, so the vectorized state remains Gaussian by conjugation. The OBDM of a pure Gaussian state is a projector, hence has occupations in $\{0,1\}$. Conversely, a projector OBDMO of a pure state forces a (generalized) Slater determinant of super-orbitals, which devectorizes into a factorized form.

	\section{Special cases}
	\label{sec:special_cases}

	We introduced above the theoretical framework for computing the  OBDMO and  how it allows to measure the amount of non-Gaussianity of an operator. In this section, we apply the framework to two important cases frequently encountered in quantum many-body physics: the time-evolution operator of a non-interacting system and a Haar random unitary. These examples are particularly insightful because they reflect contrasting notions of complexity. In non-interacting systems, the Hamiltonian -- and hence the time-evolution operator --  is diagonalized within the one-particle basis. In contrast, Haar random unitaries resemble infinite-temperature thermal states, where no simplification is expected from a one-body perspective. We demonstrate how both cases fit within the one-body density matrix framework: the time-evolution operator can be  written in a factorized form, while Haar unitaries exhibit maximal correlation entropy.

	\subsection{Non-interacting systems}

	The Hamiltonian of a non-interacting system can be put in a diagonal form by diagonalizing its Hamiltonian matrix, which constitutes an orbital rotation. In the corresponding
	basis, it has the following local form:
	\begin{equation}
		\label{eq:H_diag}
		\hat{H}_{\mathrm{free}} = \sum_{n=1}^{L} \varepsilon^{}_{n} \hat{n}^{}_{n},
	\end{equation}
	where $\varepsilon^{}_{n}$ and $\hat{n}_n$ are respectively the onsite energy and the occupation operator on the orbital $n$: each orbital is thus disconnected and all the terms
	commute with each other. The time-evolution operator reads:
	\begin{equation}
		\hat{U}(t) = \exp\left(-it\sum_{n=1}^{L} \varepsilon^{}_{n} \hat{n}^{}_{n} \right) = \prod^{L}_{n=1} e^{-it\hat{n}^{}_{n}},
	\end{equation}
	where the last equality uses the commutation relations of each term of $H$. We expand the exponential, and use the fact that $\hat{n}^{m} = \hat{n}$:
	\begin{equation}
		\label{eq:prod_U}
		\hat{U}(t) = \prod^{L}_{n=1} \left(\hat{\mathds{1}} + \hat{n}^{}_{n}\left(e^{-it\varepsilon_n}-1\right)\right) = \prod^{L}_{n=1} \hat{O}_{n}.
	\end{equation}
	The form above is quite compact, and most importantly is a factorized form. Note that the vectorized form of the operator has to be normalized as previously stated.  The vectorized form of $\hat{O}_{n}$ in its local basis  reads $|\hat{O}_{n}\vra = (1+e^{-i\varepsilon_n t} \hat{c}^{\dagger}_{n}\hat{\tilde{c}}^{}_n)|0\vra.$ Each term in the product is rotated into the following pair of orbitals (since $\hat{U}(t)$ is unitary):
	\begin{equation}
		\begin{split}
			\hh{q}^{\dagger}_{n} &= (\sqrt{A} \hat{c}^{\dagger}_n + \hat{\tilde{c}}_n), \\
			\hh{q}^{\dagger}_{L+n} &= ( \hat{c}^{\dagger}_n - \sqrt{A}\hat{\tilde{c}}_n), \\
		\end{split}
	\end{equation}
	where $A = 1-e^{-i\varepsilon_n t}$, such that $ \hh{q}^{\dagger}_{n} \hh{q}^{\dagger}_{L+n}|0\vra$ correctly reconstructs $|\hat{O}_n^{}\vra$.  Note that these orbitals must be correctly normed. Each factorized operator $\hat{O}_n$ in Eq.~\eqref{eq:prod_U} can thus be decomposed in two such kind of orbitals, which leads to:
	\begin{equation}
		|\hat{U}(t)\vra = \prod^{L}_{n=1} \hh{q}^{\dagger}_{n}\hh{q}^{\dagger}_{L+n} |0\vra = |11\cdots11\vra.
	\end{equation}
	Consequently, the operator $\hat{U}$ becomes trivial in a suitable super-orbital basis. Additionally, one can demonstrate that the
	operator $\hat{U}$ factorizes by directly evaluating matrix elements of $\Delta$, defined in Eq.~\eqref{eq:R_unitary}. Substituting Eq.~\eqref{eq:prod_U} into this expression shows
	that $\Delta^{}_{mn}=\delta_{mn}^{}\frac{e^{it\varepsilon_n}}{2}$. Using the relation between the spectra of $R$ and $\Delta$ (i.e. $\lambda^{\pm}_{\alpha} =
	\frac{1}{2}\pm|\mu_{\alpha}|$), it follows immediately that $\lambda^{\pm}_{\alpha} \in \{0,1\}$.

	The only condition imposed above is that the Hamiltonian must be quadratic in the fermionic operators. According to the Thouless theorem~\cite{Thouless_1960}, any unitary
	operator representing a one-body rotation can be expressed as the exponential of a quadratic Hamiltonian. Hence, if the spectrum of the corresponding $R$ matrix consists only of
	zeros and ones, the associated unitary operator can be interpreted as an orbital rotation. However, this factorization does not extend to the Hamiltonian itself: as evident from
	Eq.~\eqref{eq:H_diag}, a sum of local operators cannot, in general, be written as a single  factorized operator.  

	\subsection{Haar random operators}
	The case of random unitary operators is much simpler. It is possible to calculate the matrix elements $\Delta_{mn}$ on average using the common machinery of Weingarten calculus~\cite{Collins2010_Weingarten}, as shown in Appendix~\ref{app:Wg}. We find that for the time-evolution operator, these elements are all always zero, such that the spectrum of the OBDMO $R$ is fully degenerate with $\lambda=1/2$, corresponding to the worst case in which the maximal correlation entropy $S_{\mathrm{max}}=L\log(2)$ is reached, indicating maximal non-Gaussianity . In such a case, there exists no one-particle basis in which a random unitary can be simplified in terms of any of the resources that we presented. As for $\hat{U}(t)$ the eigenvalues of $\Delta$ are simply zero, we also investigate the behavior of the SNOs for a local operator which is rotated by a random unitary. In this case, we show that the eigenvalues of the matrix $\Delta$ are decaying with the system size as $2^{-L}$ which converges exponentially to a maximally correlated operator.

	\section{Results}
	\label{sec:results}

	We present the numerical study of the correlation spectrum for two specific unitary operators, the time-evolution operator $\hat{U}(t)$ and an evolved local operator  $(\hat{n}_{i}(t)-1/2)$ for two different models. The first model is the $t-V$ model, in which every site is interacting with their nearest neighbors, and the second one is the interacting
	resonant level model (IRLM) in which the interaction is localized at an edge of a tight-binding chain.  Our numerical simulations are based on a $U(1)$-symmetric MPO
	representation of the corresponding operators, and employ the time-evolving block decimation (TEBD) algorithm~\cite{Vidal_2004,Schollwock_2011} in real space with a maximum bond
	dimension $\chi=1024$, as well as a $4^{\mathrm{th}}$-order Trotter decomposition with a time step $dt=10^{-2}$ to ensure the convergence of the data up to the times shown in
	the plots. \\

	\begin{figure}[t]
		\centering
		\includegraphics[width=1.\columnwidth]{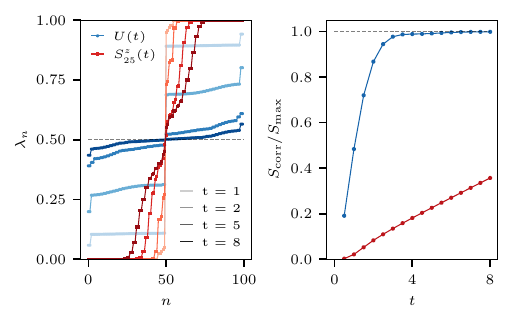}
		\caption{\textit{Left:} Spectrum of the  OBDMO of $\hat{U}(t)$ and $(\hat{n}_{25}(t)-1/2)$ for the $t$-$V$ model ($\equiv S^{z}_{25}(t)$ in the Heisenberg spin chain) with $L=50$ sites at different times.
		 The dashed line indicates $\lambda = 0.5$, corresponding to a maximally correlated orbital.
		\textit{Right:} Correlation entropy of the corresponding operators. The parameters are indicated in the text.}
		\label{fig:Heis_spectrum}
	\end{figure} 

	\subsection{\texorpdfstring{$t\text{-}V$}{tV} model}
	The first system investigated here is the $t-V$ model of spinless fermions in a one-dimensional open chain. This model is equivalent to the Heisenberg spin chain (or $XXZ$ model) through
	a Jordan-Wigner transformation~\cite{Jordan_Wigner_1928}, and is studied here as a prototypical interacting system, in which interaction (non-Gaussian terms) and hopping (Gaussian terms) compete at every site. The Hamiltonian is the following:
	\begin{equation}
		\begin{split}
			\hat{H}^{}_{\mathrm{t-V}} &= \sum_{i=1}^{L-1} \frac{\gamma}{2} \left(\hat{c}^{\dagger}_{i} \hat{c}^{}_{i+1} + \hat{c}^{\dagger}_{i+1} \hat{c}^{}_{i}\right) \\
			& + V \left(\hat{n}^{}_{i}-\frac{1}{2}\right) \left(\hat{n}^{}_{i+1} - \frac{1}{2}\right),
		\end{split}
	\end{equation}
	in which the $1/2$ subtracted to  occupation operators stems from the Heisenberg interaction $\hat{S}^{z}_{i} \hat{S}^{z}_{i+1}$ in the spin model, and enforces the particle-hole
	symmetry in the equation above.  When studying the one-body correlation spectrum of operators associated with this model, it is important to emphasize that certain
	state-specific properties, which are typically well understood, do not always exist into this framework. Conventionally, the regimes $V>0$ and $V<0$ are associated with
	anti-ferromagnetic and ferromagnetic phases, respectively, in the underlying spin model, each exhibiting distinct physical behavior. However, within our formalism, these two
	regimes are entirely equivalent.  Specifically, one can perform a basis rotation (as defined above) within the operator space $\{\mathds{1}, \hat{c}^{\dagger}_i, \hat{c}_i,
	\hat{n}_i\}^{\otimes L}$, whereby all number operators $\hat{n}^{}_{i}$ on even (or equivalently, odd) sites acquire a minus sign -- e.g., $(\hat{n}_0, \hat{n}_1, \hat{n}_2, \cdots) \rightarrow (-\hat{n}_0, \hat{n}_1, -\hat{n}_2, \cdots)$. This transformation is formally equivalent to changing $V\rightarrow-V$. As discussed in Sec.~\ref{sec:definition}, the
	Hamiltonians $\hat{H}(V)$ and $\hat{H}(-V)$, as well as the corresponding time-evolution operator $\hat{U}(t)$, are related by such an orbital rotation and therefore share the
	same correlation spectrum. A similar argument applies to the sign of the hopping parameter $\gamma$, which can be reversed by applying a rotation to every other creation (or
	equivalently annihilation) operator $\hat{c}^{\dagger}_i$ (or $\hat{c}_i$).

	In the left panel of Fig.~\ref{fig:Heis_spectrum}, we show the spectra of $R[\hat{U}(t)]$ and $R[\hat{n}^{}_i(t)-1/2]$, where $i$ is a site chosen at the center of a chain of length
	$L = 50$, for a time evolution with parameters $\gamma = V = 1$. The blue curves correspond to the unitary operator $\hat{U}(t)$ at different times, initially represented as an
	identity MPO, to which Trotter gates are sequentially applied on one side to simulate time evolution. At $t = 0$, the spectrum reflects a trivial operator, consisting solely of
	eigenvalues $0$ and $1$. As time progresses, the spectrum evolves towards that of a highly correlated operator, with most orbital occupations approaching $1/2$. The spectral gap
	$\Delta$ -- defined as the difference between the largest occupation among those close to $0$ and the smallest among those close to $1$ -- initially equal to 1, closes over time .  All orbitals become correlated similarly, lacking any notable internal structure except for the most and least occupied orbitals that are affected by the open	boundary conditions. The right panel of Fig.~\ref{fig:Heis_spectrum} presents the corresponding correlation entropy $S[\hat{U}(t)]$ as defined in Eq.~\eqref{eq:corr_ent}, normalized by
	its maximal value $L \log(2)$. This entropy rapidly saturates, indicating that no preferential single-particle basis exists for $\hat{U}(t)$, and thus its structural complexity
	cannot be reduced via orbital rotations.

	\begin{figure}[t]
		\centering
		\includegraphics[width=1.\columnwidth]{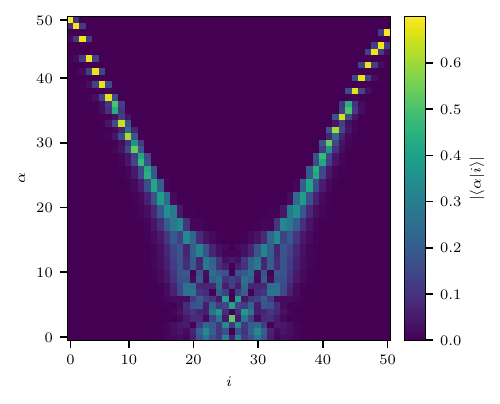}
		\caption{Spatial dispersion of the  SNOs of an evolved local operator $\hat{n}^{}_{L/2}(t)-1/2$ ($\equiv \hat{S}^{z}_{L/2}(t)$ in the spin model), for a chain of $L=50$ sites at time $t=8$. The overlap of the  SNOs with real space orbitals is defined as the absolute value of the eigenvectors of the matrix $\Delta$ as defined in Eq.~\eqref{eq:nat_orb_dec}, i.e. $|\langle \alpha| i \rangle| = |v^{}_{i\alpha}|$. The full eigenspectrum of the  OBDMO $R$ contains $2L$ orbitals, but the entire colormap consists simply of four copies of the region displayed here due to the properties of $R$ for unitary operators.} 
		\label{fig:Heis_dispersion}
	\end{figure}
	The red curves illustrate the spectrum of $R[\hat{n}^{}_{L/2}(t)-1/2]$ under the same conditions as for $\hat{U}(t)$. The initial operator is the identity at all sites except the
	center of the chain, where it is a local Pauli $S^z$ operator -- thus also a trivial operator in terms of its  SNOs. As time evolves, the spectrum deviates from the
	initial values $0$ and $1$ in a structured manner: an increasing number of orbitals acquire occupations near $1/2$, while the rest remain close to $0$ or $1$, with a smooth
	decay between the two regimes. This behavior reflects the light-cone structure characteristic of local operator spreading: at time $t$, the operator can be approximated as
	$\mathds{1} \otimes \cdots \otimes C \otimes \cdots \otimes \mathds{1}$, where $C$ denotes the non-trivial part of the operator within the light cone. If $C$ spans $c$ sites, then approximately $c$ orbitals exhibit occupations near $1/2$, while the remaining $L - c$ remain close to $0$ or $1$. The smooth decay between these regions arises because the operator is not an exact tensor product between $C$ and the outside region. This structure also governs the behavior of the correlation entropy, which grows approximately linearly in time at a significantly slower rate than that of the time evolution operator. At late times, when the support of $C$ extends across the entire system, the entropy eventually saturates near its maximal value.

	A more detailed understanding of the structure of the correlation spectrum of $\hat{n}^{}_i(t)-1/2$ can be obtained by examining the spatial dispersion of  SNOs. In Fig.~\ref{fig:Heis_dispersion}, we display the absolute values of the coefficients $v_{i\alpha}$, defined in Eq.~\eqref{eq:nat_orb_dec}, which describe the spatial profile of the  $\alpha$-th SNO $\hat{q}^{}_{\alpha}$ at time $t=8$. The orbital indices $\alpha$ are ordered by increasing proximity of their occupation numbers to $0$ or $1$, such that the most
	correlated orbitals -- those with occupations closest to 0.5 -- appear first. At this time, the light cone emerging from site $L=25$ has propagated approximately $8$ sites in both
	directions, which is reflected in the complex, delocalized structure of the leading  SNO for $\alpha \lesssim 16$. As the occupation numbers approach $0$ or $1$, the orbitals become
	increasingly localized, consistent with the exponential decay of correlations between regions inside and outside the light cone~\cite{Khemani_Huse_Nahum_2018,Xu_Swingle_2020}.
	Ultimately, each natural orbital becomes localized on a single site, in agreement with the cartoon of the operator structure discussed above, wherein sites outside the light
	cone are effectively in a tensor product state. Only the first $L$ SNOs (out of a total of $2L$) are shown in Fig.~\ref{fig:Heis_dispersion}, since the remaining orbitals are redundant for a Hermitian operator, see Eq.~\eqref{eq:nat_orb_dec}.

	We illustrated here that bulk interacting models are not likely to be simplified through one-body orbital rotations. However, the correlation spectrum of local operators as
	well as the spatial dispersion of their  SNOs exhibit footprints of underlying physical properties, which were until now investigated through different kind of
	mathematical objects such as OTOCs. In the following section, we demonstrate the power of  SNOs to reveal the hidden simplicity of some operators in specific
	models in which the interaction is localized on a few sites, namely quantum impurity models.

	\subsection{Impurity problem (IRLM)}

	The impurity model that we investigate in the following is the Interacting Resonant Level Model (IRLM), in which the spinless nature of the fermions makes numerical simulations
	more practical. The model consists in two orbitals between which fermions can hop and interact (the impurity), coupled to a large tight-binding chain of non-interacting fermions 
	(the bath). The Hamiltonian reads:
	\begin{equation}
		\begin{split}
			\hat{H}^{}_{\mathrm{IRLM}} &= U\left(\hat{n}^{}_{1} - \frac{1}{2}\right)\left(\hat{n}^{}_{2} - \frac{1}{2}\right) + V\left(\hat{c}^{\dagger}_{1}\hat{c}^{}_{2} +
			\hat{c}^{\dagger}_{2}\hat{c}^{}_{1}\right)\\ &+ \frac{\gamma}{2}\sum_{i=2}^{L-1} \left(\hat{c}^{\dagger}_{i} \hat{c}^{}_{i+1} + \hat{c}^{\dagger}_{i+1}
			\hat{c}^{}_{i}\right),
		\end{split}
	\end{equation}
	where the $1/2$ subtracted to the  occupation operators enforce particle-hole symmetry, and $\gamma/2$ is taken for the half-bandwidth of the bath to be $D=1$ for $\gamma=1$.
	Hence we work in units of $D$ (or equivalently of $\gamma$) in the following. As for the $t-V$ model presented above, the correlation spectrum is symmetric under super-orbital rotations, which in this
	model also corresponds to flipping the signs of the parameters: $U\rightarrow-U$, as well as $V\rightarrow -V$ or $\gamma\rightarrow -\gamma$ yields the same properties of the
	correlation spectrum.  In the IRLM, the ground state properties of the model with $U>0$ and $U<0$ are very different, as the Kondo temperature is exponentially smaller for
	negative values of the interaction, leading to a more correlated state~\cite{Debertolis_Florens_Snyman_2021}. Such an effect is not present here as the properties of the operator are
	related to its full spectrum, such that specific properties related to exceptional states (like the ground state) are completely diluted. In the following, we first discuss the
	properties of the time-evolution operator $\hat{U}(t)$ for different interaction strengths $0<U<D$ and $U>D$, and then we study the properties of an evolved local operator on
	the impurity site $\hat{n}_{1}(t)-1/2 \equiv \hat{S}^{z}_{1}(t)$. The numerical parameters that we choose for our simulations are the same as the ones used for the Heisenberg model. \\

	\begin{figure}[t]
		\centering
		\includegraphics[width=1.\columnwidth]{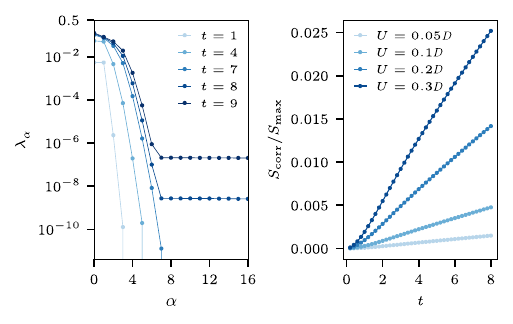}
		\caption{\textit{Left:}  Spectrum of the OBDMO for the time-evolution operator $\hat{U}(t)$ at different times for the IRLM, showing the half of orbitals the closest to $0$, the other
		half being the particle-hole symmetric $1-\lambda_\alpha$. The plateau arising after $t\geq 8$ is due to the truncation error, as the corresponding MPO has reached the
		maximum bond dimension $\chi=1024$. The $x$-axis is cut at $\alpha=16$ instead of $\alpha=2L$ for visibility, as the remaining eigenvalues are below the machine precision.
		\textit{Right:} Correlation entropy of $\hat{U}(t)$ as a function of time for different values of the interaction in the regime $U<D$, normalized by the maximal extensive
		value $S_{\mathrm{max}}=L\log{2}$. For both panels, we consider a system of $L=50$ sites and $V=0.2D$.}
		\label{fig:U_small_spectrum}
	\end{figure}

	\textit{Time-evolution operator:}
	The  spectrum of the OBDMO of the operator $\hat{U}(t)$ is shown in the left panel of Fig.~\ref{fig:U_small_spectrum} for $U = 0.3D$ in a system with $L=50$ sites. We present the spectrum at various times up to $t = 9$, beyond which a plateau appears in the spectrum around $10^{-6}$, attributable to truncation errors. The figure displays only
	the part of the spectrum near $0$, as the full spectrum is symmetric about $1$. The key observation is that, similar to the ground state of this 
	model~\cite{Debertolis_Florens_Snyman_2021,Debertolis_Snyman_Florens_2022}, the occupations of the  SNOs of the time-evolution operator decay exponentially at all times. However, this decay becomes progressively slower as time increases, eventually leading the
	spectrum to converge toward the maximal value of the correlation entropy $S_{\mathrm{max}}=L\log(2)$.
 
	The right panel of Fig.~\ref{fig:U_small_spectrum} shows the time evolution of the correlation entropy for different interaction strengths in the weak-coupling regime $U < D$. After a timescale comparable to the electronic bandwidth $D$, the correlation entropy begins to grow linearly with time. As the interaction strength increases, the entropy grows more rapidly, indicating that the operator becomes more entangled -- i.e., the occupations of a larger number of SNOs deviate more significantly from $0$ and $1$. Nonetheless, the absolute value of the correlation entropy remains well below the maximal limit, implying that the operator admits a  compact representation in the  SNO basis as we show in the following section. In contrast, its real space MPO representation used in these simulations reaches the imposed bond dimension $\chi = 1024$ already at $t = 8$, due to the much faster growth of the von Neumann entropy in that basis. In the following, we show how the rotation to  SNOs offers great compression of the considered evolved operators. 

	\begin{figure}[t]
		\centering
		\includegraphics[width=1.\columnwidth]{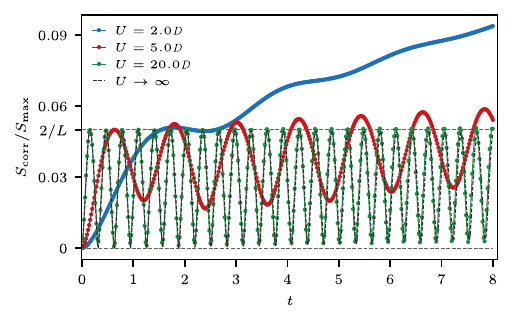}
		\caption{Correlation entropy of the time-evolution operator $\hat{U}(t)$ against time for the IRLM with $L=50$ sites, in the large-interaction regime $U>D$.  The correlation entropy is normalized by its maximal value $S_{\mathrm{max}} = L\log{2}$. The gray curve labeled by $U\to\infty$ corresponds to the analytical result at first order in $1/U$, for $U=20D$. }
		\label{fig:U_large_specrum}
	\end{figure}

	The time evolution of the correlation entropy in the large-interaction regime $U > D$ is shown in Fig.~\ref{fig:U_large_specrum}. For moderate interaction strength $U = 2D$, close to the bandwidth $D$, the correlation entropy continues to increase with time, but superimposed oscillations begin to emerge. As  the interaction $U$ gets stronger, these oscillations become more pronounced, with higher frequencies and amplitudes bounded by $2\log(2)$. This behavior can be understood by considering the limit $U \to \infty$, where the impurity becomes effectively decoupled from the bath. In this limit, the interaction at the impurity no longer influences the bath orbitals, which consequently factorize and do not contribute to the correlation entropy. The relevant degrees of freedom are then restricted to the two impurity sites, giving rise to four non-trivial  SNOs. In the decoupled limit, the occupations of these  SNOs exhibit oscillatory dynamics with distinct frequencies, governed by expressions of the form:
	\begin{equation}
		\label{eq:largeUocc}
		\lambda^{}_{\pm}(t)_{U\to\infty} = \frac{1}{2} \pm \frac{1}{2}\cos\left((V \pm \frac{U}{2})\,t\right),
	\end{equation}
	see Appendix~\ref{app:Large_U} for details on the derivation. These slightly detuned frequencies produce a correlation entropy that oscillates with two characteristic frequencies: a fast component $\omega_1 = 2/U$, and a slow component $\omega_2 = 1/V$. Additionally, this toy model indicates that spatial profiles of the SNOs oscillate between the two impurity sites. When one site is fully occupied and the other is empty, the orbitals factorize and the correlation entropy vanishes. Conversely, when both sites are half-filled, the orbitals are maximally entangled, resulting in the maximum entropy for four orbitals, $2\log(2)$. The analytical expression for the correlation entropy at first order in $1/U$ corresponds to the gray curve in Fig.~\ref{fig:U_large_specrum}, and the rescaled maximal value $2/L$ is indicated by a dashed line. The fast oscillations are correctly captured by the frequency $\omega_1$ for $U=20D$. This is not the case for the slow component: in Fig~\ref{fig:U_large_specrum}, using the simulation's parameter $V$ leads to a faster oscillating envelope, which we had to rescale here to match the data. In the toy model, the impurity is completely decoupled to the bath and is then agnostic to any bandwidth, which must play a role in the slow oscillating component related to low energies. \\

	\begin{figure}[t]
		\centering
		\includegraphics[width=1.\columnwidth]{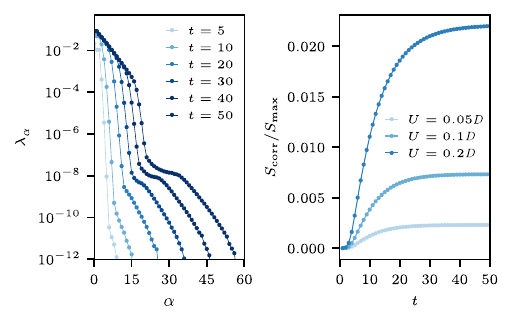}
		\caption{\textit{Left:} Occupation of the  SNOs of $\hat{n}^{}_{1}(t)-1/2$ in the IRLM for $L=100$ and $V=U=0.2D$ at different times, plotted as in Fig.~\ref{fig:U_small_spectrum}. 
		\textit{Right:} Correlation entropy of the same operator against time for different interaction strengths. For larger interaction strengths (keeping $U<D$), the correlation entropy saturates later in time at a larger value.}
		\label{fig:Sz_spectrum}
	\end{figure}

	\textit{Local operator:} We now investigate the properties of the time-evolved local operator, focusing on the $U<D$ regime, where impurity-bath correlations are not spoiled by local effects. The correlation spectrum of $\hat{n}_{1}(t) - 1/2$ is shown in the left panel of Fig.~\ref{fig:Sz_spectrum} as a function of time, for system size $L = 100$, and parameters $V = U = 0.2D$, up to time $t = 50$. The light-cone structure simplifies the MPO representation of this operator, such that no truncation effects are observed for these parameters. Nevertheless, the bond dimension keeps increasing and reaching later times is not possible.

\begin{figure}[t]
		\centering
		\includegraphics[width=1.\columnwidth]{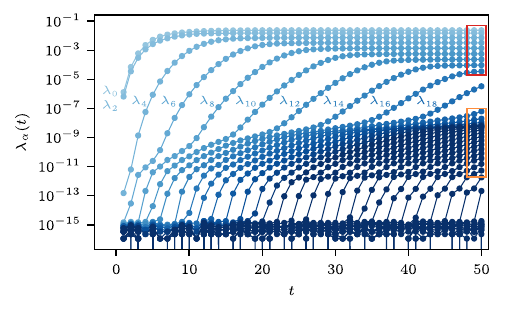}
		\caption{Time evolution of the occupations of the  SNOs of the local operator $\hat{n}_1(t)-1/2$ for $L=100$, $V=0.2D$ and $U=0.1D$. A smaller interaction strength $U$ is chosen to appreciate a faster
		exponential decay of the largest eigenvalues. The colored rectangles indicate the two groups of  SNO based on their occupation.}
		\label{fig:Evolution_Sz}
	\end{figure}

	Our first observation is that the natural orbital occupations of this operator decay exponentially as for $\hat{U}(t)$. However, the structure of the correlation spectrum is richer: we identify two distinct groups of  SNOs -- one with relatively large occupations, and a second with significantly smaller occupations with an elbow-like feature. As time progresses, orbitals from the second group transition into the first, populating the tail of the exponential decay. This behavior is illustrated in more detail in Fig.~\ref{fig:Evolution_Sz}, where the first and second groups are highlighted by red and orange rectangles, respectively.

	Fig.~\ref{fig:Evolution_Sz} presents the time evolution of the occupations of every other  SNOs, clearly showing the gradual transfer from the second group to the first. The increase in occupation of these orbitals reflects the expansion of the light-cone structure of the operator: as each new site becomes involved, it contributes a correlated orbital to the first group. Each newly included orbital exhibits an occupation that is exponentially smaller than the previous ones, with the decay rate governed by the interaction strength $U$. As with the time-evolution operator, a larger $U$ results in a slower decay, and thus a more correlated operator.

	Eventually, the gap between the two groups vanishes, signaling a saturation in the complexity of the operator. This saturation is captured in the right panel of Fig.~\ref{fig:Sz_spectrum}, which shows the evolution of the correlation entropy. The point of saturation appears to depend on the initial gap between the two orbital groups -- related to the ratio $U/V$ -- and the rate of exponential decay within the first group, which is controlled by $U/D$. \\

\begin{figure}[t]
		\centering
		\includegraphics[width=1.\columnwidth]{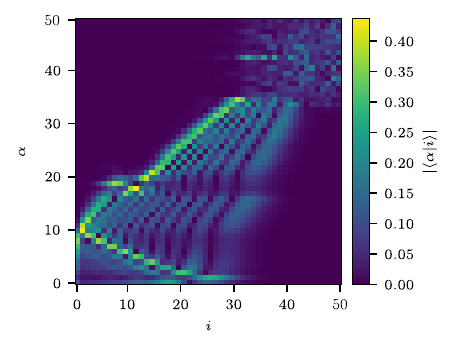}
		\caption{Spatial dispersion of the  SNOs $q^{}_{\alpha}$ of the evolve density operator on the impurity $\hat{n}^{}_{1}(t)-1/2$, for a chain of $L=100$ sites
		at time $t=30$, for $U=V=0.2D$. Only the first   $\approx 35$ eigenvalues are above machine precision, such that the eigenvectors of the  70 remaining eigenvalues are random due to spurious degeneracies. Only the $50$ first eigenvalues are thus shown for visibility.} 
		\label{fig:Sz_spatial_disp}
	\end{figure}

	The spatial distribution of the  SNOs associated to the local operator $\hat{n}_1(t)-1/2$, as depicted in Fig.~\ref{fig:Sz_spatial_disp}, reflects the hierarchical organization of orbital correlations, which decay exponentially. As previously discussed in the context of Fig.~\ref{fig:Heis_dispersion}, the most strongly correlated orbitals correspond to indices $\alpha \lesssim 30$. In contrast, for large values of $i$ and $\alpha$, the eigenvectors of the  OBDMO appear effectively random. This is a consequence of the associated eigenvalues falling below machine precision, rendering them numerically degenerate. These orbitals lie outside the light-cone and, when projected onto the real-space basis, resemble the structure observed in the exterior region of the light-cone in Fig.~\ref{fig:Heis_dispersion}. 

	The two aforementioned distinct groups of orbitals can also be identified by their spatial dispersion. The first group, composed of the most strongly correlated orbitals, exhibits a delocalized structure. As the orbital correlation strength decreases (corresponding to increasing $\alpha$), these orbitals progressively localize, eventually leading to an orbital predominantly localized at the first site. Beyond this point, the second group of orbitals -- characterized by much lower occupations emerges, displaying localization patterns around sites increasingly distant from the impurity following a linear light-cone-like propagation.  This behavior suggests distinct roles for the two orbital groups. The highly correlated, delocalized  SNOs encode the strong, shorter-ranged correlations induced by the local impurity interaction, akin to the formation of a Kondo cloud in the ground state of quantum impurity models. In contrast, the second group captures the propagation of the local perturbation through the system via nearest-neighbor hopping, representing the light-cone structure of the time-evolved operator. At long times and for sufficiently large systems, these orbitals become effectively localized on individual sites, indicating that their dynamics are no longer influenced by the impurity. This supports the observation that the correlation entropy of the local operator saturates in time, even as system size increases. The spatial extent of the most correlated orbital increases with time until it saturates at a finite length scale, which itself grows with the interaction strength $U$ (in the regime $U < D$). Both the spatial spread and the number of  SNOs in this highly correlated group increase with $U$ in this regime. The other regime $U > D$ is less interesting, as the most correlated orbitals become more localized near the impurity, and the second group of  SNOs becomes dominant.

	\subsection{Early growth of the correlation entropy}

	\begin{figure}[t]
		\centering
		\includegraphics[width=1.\columnwidth]{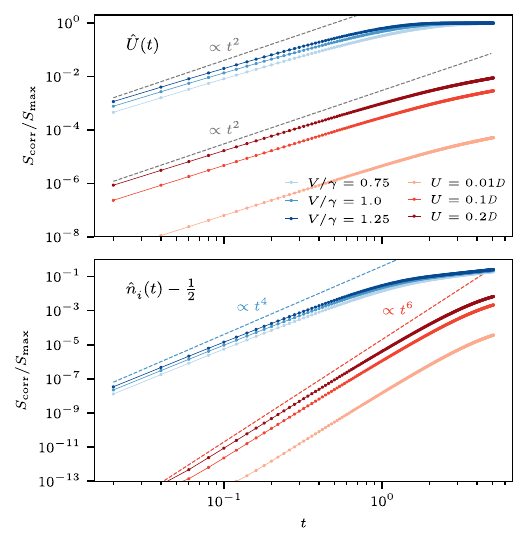}
		\caption{Correlation entropy of the time-evolution operator $\hat{U}(t)$ (upper panel) and of an evolving local operator $\hat{n}_{i}-1/2$ (lower panel), for chains of
		$L=50$ in the $t-V$ model (blue curves) and in the IRLM model (red curves). The dashed curves indicate the expected scalings from perturbation theory (see text).}
		\label{fig:Early_growth}
	\end{figure}

	Finally, we study the short time behaviour of the operators studied above for the two models presented here, the $t-V$ (or Heisenberg) model and the IRLM. At short times, the elements of the  OBDMO of these unitary operators defined in Eq.~\eqref{eq:trace_definition} can be approximated using the following form of the Baker-Hausdorff-Campell (BHC) formula:
	\begin{equation}
		e^{\hat{X}}\hat{Y}e^{-\hat{X}} = \sum_{n=0}^{\infty} \frac{1}{n!}[ \hat{X}, \hat{Y}]_p,
	\end{equation}
	where $[ \hat{X}, \hat{Y}]_p = [ \hat{X}, [\hat{X}, \hat{Y}]]_{p-1}$ and $[\hat{X}, \hat{Y}]_0=\hat{Y}$. 

	We first consider the time-evolution operator $\hat{U}(t) = e^{-i\hat{H}t}$ and compute the matrix elements of $\Delta$ to leading orders in $t$. In this framework, we define $\hat{X} = i\hat{H}t$ and $\hat{Y} = \hat{c}_m^\dagger$, while $\hat{c}_n$ remains time-independent and serves as a spatial probe for the spread of $\hat{c}_m^\dagger$ under time evolution, in a manner analogous to an OTOC.

	For the Heisenberg model, the diagonal elements of $\Delta$ are constant at leading order, with $\Delta_{mm} = 1/2$. For the off-diagonal elements, we find that $\hat{U}^\dagger \hat{c}_m^\dagger \hat{U}$ commutes with $\hat{c}_n$ up to order $p = |n - m|$ in the BCH expansion, implying $\Delta_{mn} \propto t^{|n - m|}$.  Consequently, retaining only the leading order contribution for $|n - m| = 1$ reduces $\Delta$ to a tridiagonal form. Following Eq.~\eqref{eq:R_unitary}, we compute the spectrum of $\Delta^{\dagger}\Delta$ which is also a tridiagonal matrix. Its spectrum can then be obtained analytically and is given by $\mu_\alpha = 1/2 - t^2 c_\alpha/2$, where the coefficients $c_\alpha$ encode spectral details independent of $t$. Accordingly, the spectrum of $R[\hat{U}(t)]$ is given by $\lambda_{\alpha,+} = 1 - t^2 c_\alpha$ and $\lambda_{\alpha,-} = t^2 c_\alpha$. Due to the isotropy of the Heisenberg model, all eigenvalues exhibit the same scaling behavior, leading to an extensive number of  SNOs becoming more correlated at the same time.

	In contrast, for the IRLM, the interaction anisotropy causes the diagonal elements of $\Delta$ in the bath to vanish as $t^{|m|}$ for $m > 2$, so that only the impurity contributions remain dominant at leading order. In this case, the scaling of eigenvalues remains $\lambda_{\alpha,+} = 1 - t^2 c_\alpha$ and $\lambda_{\alpha,-} = t^2 c_\alpha$, but applies only to the leading modes ($\alpha = 1, 2$), while all other eigenvalues are either $0$ or $1$. These behaviors are illustrated in the upper panel of Fig.~\ref{fig:Early_growth}, which shows the correlation entropy for both models across different parameters for a system size of $L = 50$. As predicted, the correlation entropy grows quadratically in time ($\propto t^2$) for both models. However, the amplitude is larger in the Heisenberg case, reflecting the fact that more  SNOs contribute at the same leading order.

	The case of the local operator $\hat{n}_{i}(t) - 1/2$ requires a slightly different treatment. As before, we set $\hat{X} =i\hat{H}t$, but now choose $\hat{Y} = \hat{n}_{i}$. For the $t-V$ (or Heisenberg) model, the site index $i$ is taken at the center of the chain ($i = L/2$), while for the IRLM, $i=1$ corresponds to the impurity site. In both models, the diagonal and off-diagonal elements of the  OBDMO decay faster than $1/t$ as their distance from site $i$ increases. Consequently, the dominant contributions at leading order arise from a reduced $(3 \times 3)$ sub-matrix of $\Delta$, corresponding to indices $m, n = i-1$ to $i+1$. The structural differences between the two Hamiltonians lead to distinct scaling behaviors of the spectrum of $\Delta$: in the Heisenberg model, the spectrum scales as $S_{\mathrm{corr}} \propto t^4$ (first non-vanishing terms), whereas in the IRLM it scales as $S_{\mathrm{corr}} \propto t^6$. These predictions are confirmed by numerical simulations, as shown in the bottom panel of Fig.~\ref{fig:Early_growth}.

	This highlights that the short-time scaling of the  spectrum of the OBDMO of the time-evolution operator may be insensitive to fine details of the Hamiltonian. In contrast,  local operators in the Heisenberg picture provide a more refined probe, capturing model-specific features. This sensitivity is reminiscent of the behavior of OTOC, which also involve the dynamics of local operators.

\section{MPO compression in the SNO basis}

	\begin{figure*}[t]
		\centering
		\includegraphics[width=1.\textwidth]{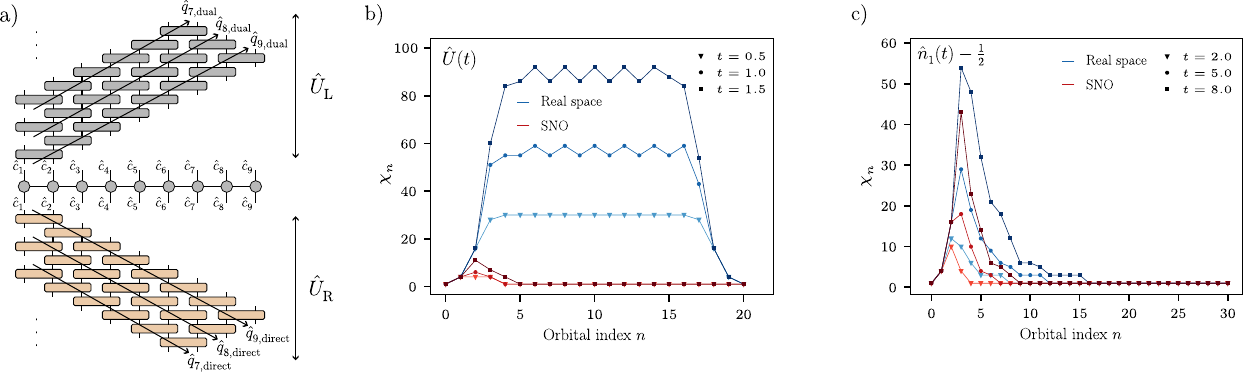}
		\caption{a) Quantum circuit rotating the MPO in the SNO basis: the right-most SNOs have the occupations closest to $0$ or $1$, and the left-most are the correlated SNOs, whose occupations are the closest to $1/2$. Each layer of local Givens rotations extract a single SNO on each side of the operator. b) Bond dimension of the time-evolution operator of the IRLM represented as an MPO in the real space and SNO bases, for $U=V=0.2D$ and $L=20$. c) Bond dimension of a local operator in the Heisenberg picture $\hat{n}^{}_{1}(t)-1/2$ in the IRLM represented as an MPO both in the real space and SNO bases, for $U=V=0.2D$ and $L=30$. }
		\label{fig:compression}
	\end{figure*}

In this section, we demonstrate how the structured non-Gaussianity of operators studied above enables a significant reduction in the complexity of the associated MPO. For impurity problems, we have shown that the occupations of the SNOs approach $0$ and $1$ exponentially, implying that orbitals whose occupations fall below machine precision are effectively factorized. In a tensor network, a factorized degree of freedom corresponds to a local tensor with bond dimension $\chi = 1$. The underlying principle of MPO compression via super orbital rotations exploits the fact that a substantial number of SNOs are factorized and can be grouped on the same side of the chain. Utilizing the properties of the eigendecomposition of the OBDMO define in Sec.~\ref{sec:definition}, each side of the operator must be rotated to a distinct set of orbitals -- the direct ($\{\hat{q}^{}_{\alpha, \text{direct}} \}$) and dual ($\{ \hat{q}^{}_{\alpha, \text{dual}} \}$) super orbital set, unless the operator is also Hermitian, in which case the left and right rotations coincide. The ordering of these orbitals in the direct and dual spaces is crucial for revealing the compressed structure of the MPO: the two orbitals associated with the same singular value of the matrix $\Delta$ must be placed at the same site, one in the direct space (physical legs pointing downward) and one in the dual space (physical legs pointing upward).

To implement the compression in practice, the orbital rotation must be applied to both sides of the MPO, which can substantially increase the entanglement. However, we employ the same techniques developed for orbital rotations of MPS~\cite{Fishman_2015,Wu_2025,Nunez_Debertolis_Florens_2025}, wherein the rotation is decomposed into a sequence of nearest-neighbor gates, as illustrated in Fig.~\ref{fig:compression}~a). Here, the local gates consist of sequences of Givens rotations, constructed to diagonalize either $\Delta\Delta^{\dagger}$ or $\Delta^{\dagger}\Delta$ for the left and right rotations, respectively. Each layer of gates extracts one orbital at the end of the chain by successively orthogonalizing each site with respect to this pair of SNOs. The direct and dual SNOs are rotated simultaneously, such that the factorized structure emerges progressively as additional orbitals are extracted. In the first step, the pair of SNOs with occupations closest to $0$ or $1$ is selected and placed at position $L$ at the end of the chain. If the occupation satisfies $\epsilon < n_\alpha < 1-\epsilon$, where $\epsilon$ denotes machine precision, the orbital factorizes and the bond dimension at this boundary reduces to $\chi = 1$. The second layer is then applied to extract the next pair of SNOs with occupations closest to $0$ or $1$, placing them at site $L-1$, and so forth. Since the application of Givens rotations modifies the current orbital basis defining the MPO, the OBDMO is rotated in the corresponding basis after each layer to determine the following appropriate rotation. Throughout this procedure, the bond dimension typically increases in the part of the MPO that is not in the SNO basis yet, before decreasing again to reach the compressed form, limiting the system sizes and time reached for demonstrating the MPO compression. Taking the full advantage of the SNOs requires the development of algorithms that do not use the real space representation of the MPO, equivalently to the \emph{few-body revealing} algorithm developed for MPS~\cite{Nunez_Debertolis_Florens_2025}.

We present the bond dimension before and after rotation for the time evolution operator $\hat{U}(t)$ of the IRLM in Fig.~\ref{fig:compression}~b) and for the local operator $\hat{n}_1^{}(t)-1/2$ in Fig.~\ref{fig:compression}~c). In real space, the orbital index corresponds to the site of the chain, whereas in the SNO basis, the MPO is constructed by ordering the SNOs according to increasing distance of their occupations from $1/2$, from left to right: with such a choice, all factorized SNOs are on the right part of the MPO and the corresponding bond dimension is thus $1$. For this numerical experiment, the MPO is evolved in real space, and a copy is rotated at each time step to compare the bond dimension of both objects.\\
For $\hat{U}(t)$, the operator is unitary but not Hermitian, necessitating distinct rotations on the left and right sides. In real space, the bond dimension remains approximately uniform along the chain and grows steadily with time, precluding simulations beyond short timescales. In contrast, the SNO basis yields a substantially compressed representation for the times considered here. As demonstrated in Fig.~\ref{fig:U_small_spectrum}, only a small number of SNOs exhibit occupations exceeding machine precision (approximately $10^{-15}$). These SNOs correspond to bond dimensions $\chi > 1$ in the MPO, while the vast majority factorize with $\chi = 1$, thereby illustrating the effectiveness of the SNO basis in revealing factorized structures within tensor network representations.\\
For the local operator $\hat{n}_1^{}(t)-1/2$, which is both unitary and Hermitian, the left and right rotations coincide. In real space, the bond dimension along the chain exhibits a light-cone structure, with sites beyond the light-cone front being factorized. In the SNO basis, the operator admits a compressed representation even within the light-cone region, resulting in a larger number of factorized super orbitals and overall reduced bond dimensions. Furthermore, even within the light-cone, the ordering based on SNO occupations appears to play a significant role, suggesting that the chosen ordering is also beneficial in the entangled portion of the MPO. The large bond dimension in real space prevents us from performing the rotation at later times for the system sizes considered. However, a striking property of the non-Gaussianity of local operators in the Heisenberg picture, as shown in Fig.~\ref{fig:Sz_spectrum}, is its saturation over time: the SNO occupations cease to evolve. Consequently, the complexity of the MPO is expected to saturate at long times for such local operators.\\

\emph{Multiplying rotated MPOs:} In order to compute physical quantities such as OTOCs, we might want to compute traces of different operators. These operators might have compressed representations in different bases, and there are two main strategies to tackle such a product. 

(i): the first operator is evolved in the Heisenberg picture and compressed in some basis $U^{}$, and the second one is local (the typical case of an OTOC). We write $\hat{\Omega}(U)$ the Fock space representation of the $(L\times L)$ matrix $U$, such that $\hat{O} = \hat{\Omega}(U_{L}^{\dagger}) \hat{\Tilde{O}} \hat{\Omega}(U_R^{})$, with $\hat{\Tilde{O}}$ the  compressed MPO in its SNO basis. We want $\mathrm{Tr}(\hat{\Omega}(U_{R}^{\dagger}) \hat{\Tilde{O}}^{\dagger} \hat{\Omega}(U_L^{}) \hat{O}_j \hat{\Omega}(U_{L}^{\dagger}) \hat{\Tilde{O}} \hat{\Omega}(U_R^{})\hat{O}_j)$. One way is to decompose $\hat{O}_j$ in the corresponding  left or right basis, requiring $L^{|{O}_j|}$ traces, where $|\hat{O}_j|$ is the number of (distinct) fermionic operators in $\hat{O}_j$. The other way is to localize the modes on which $\hat{O}_{j}$ act directly in $\hat{O}(t)$: one layer of Givens staircase is necessary for each mode, so the MPO bond dimension can increase at most of a factor $2^{|\hat{O}_j|}$.

(ii): both operators are evolved in the Heisenberg picture and compressed in different bases $U_1$ and $U_2$: $\mathrm{Tr}(\hat{\Omega}(U_{1R}^{\dagger}) \hat{\Tilde{O}}_1^{\dagger} \hat{\Omega}(U_{1L}^{}) \hat{\Omega}(U_{2L}^{\dagger}) \hat{\Tilde{O}}_2 \hat{\Omega}(U_{2R}^{}))$. We can combine the rotation from both operators as follows: $\mathrm{Tr}(\hat{\Omega}(U_{2R}^{} U_{1R}^{\dagger}) \hat{\Tilde{O}}_1^{\dagger} \hat{\Omega}(U_{1L}^{} U_{2L}^{\dagger}) \hat{\Tilde{O}}_2)$. From this we have to absorb the product of the rotations into $\hat{\Tilde{O}}_1$ or $\hat{\Tilde{O}}_2$, as a sequence of Givens rotations. The complexity of the resulted objects will depend on \emph{how far} from each other are $U_1$ and $U_2$: if they don't differ too much, $U_{2}^{} U_{1}^{\dagger} \approx \mathds{1}$ and the increase of bond dimension of the rotated object is limited.

	\section{Conclusion and outlook}
	\label{sec:discussion}
	In this work, we have introduced the concept of  SNOs associated with many-body operators, defined as the eigenvectors of the one-body super-density matrix constructed from a vectorized operator. To illustrate the meaning of these orbitals in the super-Hilbert space of $2L$ particles, we analyzed several simple cases. We demonstrated that a suitable super-orbital rotation can render an operator trivial in the MPO representation for non-interacting systems, while no such simplification is possible for random operators. We introduced a measure of non-Gaussianity for operators and discuss its properties related to the occupations of SNOs. Additionally, we showed that for unitary operators, a super-orbital rotation corresponds to a standard basis change that preserves the meaning of the operator in the physical Hilbert space of $L$ particles. The measure of non-Gaussianity holds for mixed states due to the double-copy structure of the OBDMO, and the corresponding orbital rotations require to work with the vectorized density matrix in the doubled space equipped with super-operators obeying canonical anticommutation relations -- computing the spectrum does not require it, but implementing the rotation does, because SNOs mix direct and dual sectors -- which we leave for future work.

	We then applied this framework to characterize the  SNOs of both the time-evolution operator and a time-evolved local operator in two interacting models: the $t-V$ model and a quantum impurity model. In the impurity model, the  SNOs reveal a compressed structure in both operators, evident through the exponential decay of their occupation numbers. Using the correlation entropy, we observed that the complexity or non-Gaussianity of a local operator initially located on the impurity site saturates over time, with the saturation timescale increasing with interaction strength. We have demonstrated numerically that rotating operators into the SNO basis with appropriate ordering yields substantial compression of the MPO. The development of an algorithm operating entirely within the compressed basis would enable the study of larger systems and longer timescales, beyond the reach of current real-space tensor network methods.

	Future directions include the development of the few-body revealing algorithm for operators, analogous to that introduced for quantum states in Ref.~\cite{Nunez_Debertolis_Florens_2025}, to efficiently simulate OTOCs in quantum impurity models at long times and for large systems, in order to test the Lieb-Robinson bound in these systems. Such a tool would validate and extend the present findings, shedding light on the scrambling behavior in these systems.  Finally, this work opens up new perspectives for studying orbital correlations in operator space across different physical contexts -- for example, in many-body localized systems, where natural orbitals associated with quantum states have shown intriguing properties~\cite{Bera_Schomerus_Meisner_Bardarson_2015,Bera_Martynec_Schomerus_Meisner_Bardarson_2017,Buijsman_Gritsev_Cheianov_2018,Hopjan_Meisner_2020}. Another promising avenue is the investigation of  SNOs in mixed-state density matrices represented as MPOs, aiming to understand the impact of interaction with the environment and to provide a one-body framework for analyzing the loss of correlations due to dissipation. \\

	\textit{Acknowledgement}. The author thank Serge Florens, Yuriel N\'u\~nez-Fern\'andez, Xhek Turkeshi and Frank Pollmann for interesting discussions. The author was supported by the Deutsche Forschungsgemeinschaft through the cluster of excellence ML4Q (EXC 2004, project-id 390534769) and by the Deutsche Forschungs- gemeinschaft through CRC 1639 NuMeriQS (project-id 511713970).

	\bibliography{Biblio.bib}

\begin{thebibliography}{81}%
\makeatletter
\providecommand \@ifxundefined [1]{%
 \@ifx{#1\undefined}
}%
\providecommand \@ifnum [1]{%
 \ifnum #1\expandafter \@firstoftwo
 \else \expandafter \@secondoftwo
 \fi
}%
\providecommand \@ifx [1]{%
 \ifx #1\expandafter \@firstoftwo
 \else \expandafter \@secondoftwo
 \fi
}%
\providecommand \natexlab [1]{#1}%
\providecommand \enquote  [1]{``#1''}%
\providecommand \bibnamefont  [1]{#1}%
\providecommand \bibfnamefont [1]{#1}%
\providecommand \citenamefont [1]{#1}%
\providecommand \href@noop [0]{\@secondoftwo}%
\providecommand \href [0]{\begingroup \@sanitize@url \@href}%
\providecommand \@href[1]{\@@startlink{#1}\@@href}%
\providecommand \@@href[1]{\endgroup#1\@@endlink}%
\providecommand \@sanitize@url [0]{\catcode `\\12\catcode `\$12\catcode
  `\&12\catcode `\#12\catcode `\^12\catcode `\_12\catcode `\%12\relax}%
\providecommand \@@startlink[1]{}%
\providecommand \@@endlink[0]{}%
\providecommand \url  [0]{\begingroup\@sanitize@url \@url }%
\providecommand \@url [1]{\endgroup\@href {#1}{\urlprefix }}%
\providecommand \urlprefix  [0]{URL }%
\providecommand \Eprint [0]{\href }%
\providecommand \doibase [0]{https://doi.org/}%
\providecommand \selectlanguage [0]{\@gobble}%
\providecommand \bibinfo  [0]{\@secondoftwo}%
\providecommand \bibfield  [0]{\@secondoftwo}%
\providecommand \translation [1]{[#1]}%
\providecommand \BibitemOpen [0]{}%
\providecommand \bibitemStop [0]{}%
\providecommand \bibitemNoStop [0]{.\EOS\space}%
\providecommand \EOS [0]{\spacefactor3000\relax}%
\providecommand \BibitemShut  [1]{\csname bibitem#1\endcsname}%
\let\auto@bib@innerbib\@empty
\bibitem [{\citenamefont {Keimer}\ \emph {et~al.}(2015)\citenamefont {Keimer},
  \citenamefont {Kivelson}, \citenamefont {Norman}, \citenamefont {Uchida},\
  and\ \citenamefont {Zaanen}}]{Keimer_Kivelson_Norman_Uchida_Zaanen_2015}%
  \BibitemOpen
  \bibfield  {author} {\bibinfo {author} {\bibfnamefont {B.}~\bibnamefont
  {Keimer}}, \bibinfo {author} {\bibfnamefont {S.}~\bibnamefont {Kivelson}},
  \bibinfo {author} {\bibfnamefont {M.}~\bibnamefont {Norman}}, \bibinfo
  {author} {\bibfnamefont {S.}~\bibnamefont {Uchida}},\ and\ \bibinfo {author}
  {\bibfnamefont {J.}~\bibnamefont {Zaanen}},\ }\bibfield  {title} {\bibinfo
  {title} {From quantum matter to high-temperature superconductivity in copper
  oxides},\ }\href {https://doi.org/10.1038/nature14165} {\bibfield  {journal}
  {\bibinfo  {journal} {Nature}\ }\textbf {\bibinfo {volume} {518}},\ \bibinfo
  {pages} {179–186} (\bibinfo {year} {2015})}\BibitemShut {NoStop}%
\bibitem [{\citenamefont {Savary}\ and\ \citenamefont
  {Balents}(2016)}]{Savary_2017}%
  \BibitemOpen
  \bibfield  {author} {\bibinfo {author} {\bibfnamefont {L.}~\bibnamefont
  {Savary}}\ and\ \bibinfo {author} {\bibfnamefont {L.}~\bibnamefont
  {Balents}},\ }\bibfield  {title} {\bibinfo {title} {Quantum spin liquids: a
  review},\ }\href {https://doi.org/10.1088/0034-4885/80/1/016502} {\bibfield
  {journal} {\bibinfo  {journal} {Reports on Progress in Physics}\ }\textbf
  {\bibinfo {volume} {80}},\ \bibinfo {pages} {016502} (\bibinfo {year}
  {2016})}\BibitemShut {NoStop}%
\bibitem [{\citenamefont {Nandkishore}\ and\ \citenamefont
  {Huse}(2015)}]{Nandkishore_Huse_2015}%
  \BibitemOpen
  \bibfield  {author} {\bibinfo {author} {\bibfnamefont {R.}~\bibnamefont
  {Nandkishore}}\ and\ \bibinfo {author} {\bibfnamefont {D.~A.}\ \bibnamefont
  {Huse}},\ }\bibfield  {title} {\bibinfo {title} {Many-body localization and
  thermalization in quantum statistical mechanics},\ }\href
  {https://doi.org/https://doi.org/10.1146/annurev-conmatphys-031214-014726}
  {\bibfield  {journal} {\bibinfo  {journal} {Annual Review of Condensed Matter
  Physics}\ }\textbf {\bibinfo {volume} {6}},\ \bibinfo {pages} {15} (\bibinfo
  {year} {2015})}\BibitemShut {NoStop}%
\bibitem [{\citenamefont {Abanin}\ \emph {et~al.}(2019)\citenamefont {Abanin},
  \citenamefont {Altman}, \citenamefont {Bloch},\ and\ \citenamefont
  {Serbyn}}]{Abanin_Altman_Bloch_Serbyn_2019}%
  \BibitemOpen
  \bibfield  {author} {\bibinfo {author} {\bibfnamefont {D.~A.}\ \bibnamefont
  {Abanin}}, \bibinfo {author} {\bibfnamefont {E.}~\bibnamefont {Altman}},
  \bibinfo {author} {\bibfnamefont {I.}~\bibnamefont {Bloch}},\ and\ \bibinfo
  {author} {\bibfnamefont {M.}~\bibnamefont {Serbyn}},\ }\bibfield  {title}
  {\bibinfo {title} {Colloquium: Many-body localization, thermalization, and
  entanglement},\ }\href {https://doi.org/10.1103/RevModPhys.91.021001}
  {\bibfield  {journal} {\bibinfo  {journal} {Rev. Mod. Phys.}\ }\textbf
  {\bibinfo {volume} {91}},\ \bibinfo {pages} {021001} (\bibinfo {year}
  {2019})}\BibitemShut {NoStop}%
\bibitem [{\citenamefont {Lanczos}(1952)}]{Lanczos1952SolutionOS}%
  \BibitemOpen
  \bibfield  {author} {\bibinfo {author} {\bibfnamefont {C.}~\bibnamefont
  {Lanczos}},\ }\bibfield  {title} {\bibinfo {title} {Solution of systems of
  linear equations by minimized iterations},\ }\href
  {https://api.semanticscholar.org/CorpusID:7484650} {\bibfield  {journal}
  {\bibinfo  {journal} {Journal of research of the National Bureau of
  Standards}\ }\textbf {\bibinfo {volume} {49}},\ \bibinfo {pages} {33}
  (\bibinfo {year} {1952})}\BibitemShut {NoStop}%
\bibitem [{\citenamefont {Benner}\ and\ \citenamefont
  {Fa{\ss}bender}(1997)}]{Benner_fassbender_1997}%
  \BibitemOpen
  \bibfield  {author} {\bibinfo {author} {\bibfnamefont {P.}~\bibnamefont
  {Benner}}\ and\ \bibinfo {author} {\bibfnamefont {H.}~\bibnamefont
  {Fa{\ss}bender}},\ }\bibfield  {title} {\bibinfo {title} {An implicitly
  restarted symplectic lanczos method for the hamiltonian eigenvalue problem},\
  }\href {https://doi.org/https://doi.org/10.1016/S0024-3795(96)00524-1}
  {\bibfield  {journal} {\bibinfo  {journal} {Linear Algebra and its
  Applications}\ }\textbf {\bibinfo {volume} {263}},\ \bibinfo {pages} {75}
  (\bibinfo {year} {1997})}\BibitemShut {NoStop}%
\bibitem [{\citenamefont {Sandvik}\ and\ \citenamefont
  {Kurkij\"arvi}(1991)}]{Sandvik_Kurkijarvi_1991}%
  \BibitemOpen
  \bibfield  {author} {\bibinfo {author} {\bibfnamefont {A.~W.}\ \bibnamefont
  {Sandvik}}\ and\ \bibinfo {author} {\bibfnamefont {J.}~\bibnamefont
  {Kurkij\"arvi}},\ }\bibfield  {title} {\bibinfo {title} {Quantum monte carlo
  simulation method for spin systems},\ }\href
  {https://doi.org/10.1103/PhysRevB.43.5950} {\bibfield  {journal} {\bibinfo
  {journal} {Phys. Rev. B}\ }\textbf {\bibinfo {volume} {43}},\ \bibinfo
  {pages} {5950} (\bibinfo {year} {1991})}\BibitemShut {NoStop}%
\bibitem [{\citenamefont {Foulkes}\ \emph {et~al.}(2001)\citenamefont
  {Foulkes}, \citenamefont {Mitas}, \citenamefont {Needs},\ and\ \citenamefont
  {Rajagopal}}]{Foulkes_Mitas_Needs_Rajagopal_2001}%
  \BibitemOpen
  \bibfield  {author} {\bibinfo {author} {\bibfnamefont {W.~M.~C.}\
  \bibnamefont {Foulkes}}, \bibinfo {author} {\bibfnamefont {L.}~\bibnamefont
  {Mitas}}, \bibinfo {author} {\bibfnamefont {R.~J.}\ \bibnamefont {Needs}},\
  and\ \bibinfo {author} {\bibfnamefont {G.}~\bibnamefont {Rajagopal}},\
  }\bibfield  {title} {\bibinfo {title} {Quantum monte carlo simulations of
  solids},\ }\href {https://doi.org/10.1103/RevModPhys.73.33} {\bibfield
  {journal} {\bibinfo  {journal} {Rev. Mod. Phys.}\ }\textbf {\bibinfo {volume}
  {73}},\ \bibinfo {pages} {33} (\bibinfo {year} {2001})}\BibitemShut {NoStop}%
\bibitem [{\citenamefont {Schollw\"ock}(2011)}]{Schollwock_2011}%
  \BibitemOpen
  \bibfield  {author} {\bibinfo {author} {\bibfnamefont {U.}~\bibnamefont
  {Schollw\"ock}},\ }\bibfield  {title} {\bibinfo {title} {The density-matrix
  renormalization group in the age of matrix product states},\ }\href
  {https://doi.org/https://doi.org/10.1016/j.aop.2010.09.012} {\bibfield
  {journal} {\bibinfo  {journal} {Annals of Physics}\ }\textbf {\bibinfo
  {volume} {326}},\ \bibinfo {pages} {96} (\bibinfo {year} {2011})}\BibitemShut
  {NoStop}%
\bibitem [{\citenamefont {Bañuls}(2023)}]{Banuls_2023}%
  \BibitemOpen
  \bibfield  {author} {\bibinfo {author} {\bibfnamefont {M.~C.}\ \bibnamefont
  {Bañuls}},\ }\bibfield  {title} {\bibinfo {title} {Tensor network
  algorithms: A route map},\ }\href
  {https://doi.org/https://doi.org/10.1146/annurev-conmatphys-040721-022705}
  {\bibfield  {journal} {\bibinfo  {journal} {Annual Review of Condensed Matter
  Physics}\ }\textbf {\bibinfo {volume} {14}},\ \bibinfo {pages} {173}
  (\bibinfo {year} {2023})}\BibitemShut {NoStop}%
\bibitem [{\citenamefont {White}(1992)}]{White_92}%
  \BibitemOpen
  \bibfield  {author} {\bibinfo {author} {\bibfnamefont {S.~R.}\ \bibnamefont
  {White}},\ }\bibfield  {title} {\bibinfo {title} {Density matrix formulation
  for quantum renormalization groups},\ }\href
  {https://doi.org/10.1103/PhysRevLett.69.2863} {\bibfield  {journal} {\bibinfo
   {journal} {Phys. Rev. Lett.}\ }\textbf {\bibinfo {volume} {69}},\ \bibinfo
  {pages} {2863} (\bibinfo {year} {1992})}\BibitemShut {NoStop}%
\bibitem [{\citenamefont {\"Ostlund}\ and\ \citenamefont
  {Rommer}(1995)}]{Ostlund_Rommer_1995}%
  \BibitemOpen
  \bibfield  {author} {\bibinfo {author} {\bibfnamefont {S.}~\bibnamefont
  {\"Ostlund}}\ and\ \bibinfo {author} {\bibfnamefont {S.}~\bibnamefont
  {Rommer}},\ }\bibfield  {title} {\bibinfo {title} {Thermodynamic limit of
  density matrix renormalization},\ }\href
  {https://doi.org/10.1103/PhysRevLett.75.3537} {\bibfield  {journal} {\bibinfo
   {journal} {Phys. Rev. Lett.}\ }\textbf {\bibinfo {volume} {75}},\ \bibinfo
  {pages} {3537} (\bibinfo {year} {1995})}\BibitemShut {NoStop}%
\bibitem [{\citenamefont {Larkin}\ and\ \citenamefont
  {Ovchinnikov}(1969)}]{Larkin_1969}%
  \BibitemOpen
  \bibfield  {author} {\bibinfo {author} {\bibfnamefont {A.~I.}\ \bibnamefont
  {Larkin}}\ and\ \bibinfo {author} {\bibfnamefont {Y.~N.}\ \bibnamefont
  {Ovchinnikov}},\ }\bibfield  {title} {\bibinfo {title} {Quasiclassical method
  in the theory of superconductivity},\ }\href
  {https://api.semanticscholar.org/CorpusID:117608877} {\bibfield  {journal}
  {\bibinfo  {journal} {Journal of Experimental and Theoretical Physics}\ }
  (\bibinfo {year} {1969})}\BibitemShut {NoStop}%
\bibitem [{\citenamefont {Maldacena}\ \emph {et~al.}(2016)\citenamefont
  {Maldacena}, \citenamefont {Shenker},\ and\ \citenamefont
  {Stanford}}]{Maldacena_Shenker_Stanford_2016}%
  \BibitemOpen
  \bibfield  {author} {\bibinfo {author} {\bibfnamefont {J.}~\bibnamefont
  {Maldacena}}, \bibinfo {author} {\bibfnamefont {S.~H.}\ \bibnamefont
  {Shenker}},\ and\ \bibinfo {author} {\bibfnamefont {D.}~\bibnamefont
  {Stanford}},\ }\bibfield  {title} {\bibinfo {title} {A bound on chaos},\
  }\href {https://doi.org/10.1007/JHEP08(2016)106} {\bibfield  {journal}
  {\bibinfo  {journal} {Journal of High Energy Physics}\ }\textbf {\bibinfo
  {volume} {2016}},\ \bibinfo {pages} {106} (\bibinfo {year}
  {2016})}\BibitemShut {NoStop}%
\bibitem [{\citenamefont {Bohrdt}\ \emph {et~al.}(2017)\citenamefont {Bohrdt},
  \citenamefont {Mendl}, \citenamefont {Endres},\ and\ \citenamefont
  {Knap}}]{Bohrdt_2017}%
  \BibitemOpen
  \bibfield  {author} {\bibinfo {author} {\bibfnamefont {A.}~\bibnamefont
  {Bohrdt}}, \bibinfo {author} {\bibfnamefont {C.~B.}\ \bibnamefont {Mendl}},
  \bibinfo {author} {\bibfnamefont {M.}~\bibnamefont {Endres}},\ and\ \bibinfo
  {author} {\bibfnamefont {M.}~\bibnamefont {Knap}},\ }\bibfield  {title}
  {\bibinfo {title} {Scrambling and thermalization in a diffusive quantum
  many-body system},\ }\href {https://doi.org/10.1088/1367-2630/aa719b}
  {\bibfield  {journal} {\bibinfo  {journal} {New Journal of Physics}\ }\textbf
  {\bibinfo {volume} {19}},\ \bibinfo {pages} {063001} (\bibinfo {year}
  {2017})}\BibitemShut {NoStop}%
\bibitem [{\citenamefont {H\'emery}\ \emph {et~al.}(2019)\citenamefont
  {H\'emery}, \citenamefont {Pollmann},\ and\ \citenamefont
  {Luitz}}]{Hemery_Pollmann_Luitz_2019}%
  \BibitemOpen
  \bibfield  {author} {\bibinfo {author} {\bibfnamefont {K.}~\bibnamefont
  {H\'emery}}, \bibinfo {author} {\bibfnamefont {F.}~\bibnamefont {Pollmann}},\
  and\ \bibinfo {author} {\bibfnamefont {D.~J.}\ \bibnamefont {Luitz}},\
  }\bibfield  {title} {\bibinfo {title} {Matrix product states approaches to
  operator spreading in ergodic quantum systems},\ }\href
  {https://doi.org/10.1103/PhysRevB.100.104303} {\bibfield  {journal} {\bibinfo
   {journal} {Phys. Rev. B}\ }\textbf {\bibinfo {volume} {100}},\ \bibinfo
  {pages} {104303} (\bibinfo {year} {2019})}\BibitemShut {NoStop}%
\bibitem [{\citenamefont {Xu}\ and\ \citenamefont
  {Swingle}(2020)}]{Xu_Swingle_2020}%
  \BibitemOpen
  \bibfield  {author} {\bibinfo {author} {\bibfnamefont {S.}~\bibnamefont
  {Xu}}\ and\ \bibinfo {author} {\bibfnamefont {B.}~\bibnamefont {Swingle}},\
  }\bibfield  {title} {\bibinfo {title} {Accessing scrambling using matrix
  product operators},\ }\href {https://doi.org/10.1038/s41567-019-0712-4}
  {\bibfield  {journal} {\bibinfo  {journal} {Nature Physics}\ }\textbf
  {\bibinfo {volume} {16}},\ \bibinfo {pages} {199–204} (\bibinfo {year}
  {2020})}\BibitemShut {NoStop}%
\bibitem [{\citenamefont {Gisti}\ \emph {et~al.}(2025)\citenamefont {Gisti},
  \citenamefont {Luitz},\ and\ \citenamefont
  {Debertolis}}]{Gisti_Luitz_Debertolis_2025}%
  \BibitemOpen
  \bibfield  {author} {\bibinfo {author} {\bibfnamefont {M.}~\bibnamefont
  {Gisti}}, \bibinfo {author} {\bibfnamefont {D.~J.}\ \bibnamefont {Luitz}},\
  and\ \bibinfo {author} {\bibfnamefont {M.}~\bibnamefont {Debertolis}},\
  }\bibfield  {title} {\bibinfo {title} {Symmetry resolved out-of-time-order
  correlators of heisenberg spin chains using projected matrix product
  operators},\ }\href {https://arxiv.org/abs/2503.20327} {\bibfield  {journal}
  {\bibinfo  {journal} {arXiv:2503.20327}\ } (\bibinfo {year}
  {2025})}\BibitemShut {NoStop}%
\bibitem [{\citenamefont {Calabrese}\ and\ \citenamefont
  {Cardy}(2005)}]{Calabrese_2005}%
  \BibitemOpen
  \bibfield  {author} {\bibinfo {author} {\bibfnamefont {P.}~\bibnamefont
  {Calabrese}}\ and\ \bibinfo {author} {\bibfnamefont {J.}~\bibnamefont
  {Cardy}},\ }\bibfield  {title} {\bibinfo {title} {Evolution of entanglement
  entropy in one-dimensional systems},\ }\href
  {https://doi.org/10.1088/1742-5468/2005/04/P04010} {\bibfield  {journal}
  {\bibinfo  {journal} {Journal of Statistical Mechanics: Theory and
  Experiment}\ }\textbf {\bibinfo {volume} {2005}},\ \bibinfo {pages} {P04010}
  (\bibinfo {year} {2005})}\BibitemShut {NoStop}%
\bibitem [{\citenamefont {Fagotti}\ and\ \citenamefont
  {Calabrese}(2008)}]{Fagotti_Calabrese_2008}%
  \BibitemOpen
  \bibfield  {author} {\bibinfo {author} {\bibfnamefont {M.}~\bibnamefont
  {Fagotti}}\ and\ \bibinfo {author} {\bibfnamefont {P.}~\bibnamefont
  {Calabrese}},\ }\bibfield  {title} {\bibinfo {title} {Evolution of
  entanglement entropy following a quantum quench: Analytic results for the
  $xy$ chain in a transverse magnetic field},\ }\href
  {https://doi.org/10.1103/PhysRevA.78.010306} {\bibfield  {journal} {\bibinfo
  {journal} {Phys. Rev. A}\ }\textbf {\bibinfo {volume} {78}},\ \bibinfo
  {pages} {010306} (\bibinfo {year} {2008})}\BibitemShut {NoStop}%
\bibitem [{\citenamefont {Eisert}\ \emph {et~al.}(2010)\citenamefont {Eisert},
  \citenamefont {Cramer},\ and\ \citenamefont
  {Plenio}}]{Eisert_Cramer_Plenio_2010}%
  \BibitemOpen
  \bibfield  {author} {\bibinfo {author} {\bibfnamefont {J.}~\bibnamefont
  {Eisert}}, \bibinfo {author} {\bibfnamefont {M.}~\bibnamefont {Cramer}},\
  and\ \bibinfo {author} {\bibfnamefont {M.~B.}\ \bibnamefont {Plenio}},\
  }\bibfield  {title} {\bibinfo {title} {Colloquium: Area laws for the
  entanglement entropy},\ }\href {https://doi.org/10.1103/RevModPhys.82.277}
  {\bibfield  {journal} {\bibinfo  {journal} {Rev. Mod. Phys.}\ }\textbf
  {\bibinfo {volume} {82}},\ \bibinfo {pages} {277} (\bibinfo {year}
  {2010})}\BibitemShut {NoStop}%
\bibitem [{\citenamefont {Alba}(2025)}]{Alba_2024}%
  \BibitemOpen
  \bibfield  {author} {\bibinfo {author} {\bibfnamefont {V.}~\bibnamefont
  {Alba}},\ }\bibfield  {title} {\bibinfo {title} {More on the operator space
  entanglement (ose): Rényi ose, revivals, and integrability breaking},\
  }\bibfield  {journal} {\bibinfo  {journal} {Journal of Physics A:
  Mathematical and Theoretical}\ }\href
  {https://doi.org/10.1088/1751-8121/adc9e6} {10.1088/1751-8121/adc9e6}
  (\bibinfo {year} {2025})\BibitemShut {NoStop}%
\bibitem [{\citenamefont {Murciano}\ \emph {et~al.}(2024)\citenamefont
  {Murciano}, \citenamefont {Dubail},\ and\ \citenamefont
  {Calabrese}}]{Murciano_Dubail_Calabrese_2024}%
  \BibitemOpen
  \bibfield  {author} {\bibinfo {author} {\bibfnamefont {S.}~\bibnamefont
  {Murciano}}, \bibinfo {author} {\bibfnamefont {J.}~\bibnamefont {Dubail}},\
  and\ \bibinfo {author} {\bibfnamefont {P.}~\bibnamefont {Calabrese}},\
  }\bibfield  {title} {\bibinfo {title} {More on symmetry resolved operator
  entanglement},\ }\href {https://doi.org/10.1088/1751-8121/ad30d1} {\bibfield
  {journal} {\bibinfo  {journal} {Journal of Physics A: Mathematical and
  Theoretical}\ }\textbf {\bibinfo {volume} {57}},\ \bibinfo {pages} {145002}
  (\bibinfo {year} {2024})}\BibitemShut {NoStop}%
\bibitem [{\citenamefont {Núñez-Fernández}\ \emph
  {et~al.}(2025)\citenamefont {Núñez-Fernández}, \citenamefont
  {Debertolis},\ and\ \citenamefont {Florens}}]{Nunez_Debertolis_Florens_2025}%
  \BibitemOpen
  \bibfield  {author} {\bibinfo {author} {\bibfnamefont {Y.}~\bibnamefont
  {Núñez-Fernández}}, \bibinfo {author} {\bibfnamefont {M.}~\bibnamefont
  {Debertolis}},\ and\ \bibinfo {author} {\bibfnamefont {S.}~\bibnamefont
  {Florens}},\ }\bibfield  {title} {\bibinfo {title} {Resolving space-time
  structures of quantum impurities with a numerically-exact algorithm using
  few-body revealing},\ }\href {https://arxiv.org/abs/2503.13706} {\bibfield
  {journal} {\bibinfo  {journal} {arXiv:2503.13706}\ } (\bibinfo {year}
  {2025})}\BibitemShut {NoStop}%
\bibitem [{\citenamefont {Ashida}\ \emph
  {et~al.}(2018{\natexlab{a}})\citenamefont {Ashida}, \citenamefont {Shi},
  \citenamefont {Bañuls}, \citenamefont {Cirac},\ and\ \citenamefont
  {Demler}}]{Ashida_2018}%
  \BibitemOpen
  \bibfield  {author} {\bibinfo {author} {\bibfnamefont {Y.}~\bibnamefont
  {Ashida}}, \bibinfo {author} {\bibfnamefont {T.}~\bibnamefont {Shi}},
  \bibinfo {author} {\bibfnamefont {M.~C.}\ \bibnamefont {Bañuls}}, \bibinfo
  {author} {\bibfnamefont {J.~I.}\ \bibnamefont {Cirac}},\ and\ \bibinfo
  {author} {\bibfnamefont {E.}~\bibnamefont {Demler}},\ }\bibfield  {title}
  {\bibinfo {title} {Solving quantum impurity problems in and out of
  equilibrium with the variational approach},\ }\bibfield  {journal} {\bibinfo
  {journal} {Physical Review Letters}\ }\textbf {\bibinfo {volume} {121}},\
  \href {https://doi.org/10.1103/physrevlett.121.026805}
  {10.1103/physrevlett.121.026805} (\bibinfo {year}
  {2018}{\natexlab{a}})\BibitemShut {NoStop}%
\bibitem [{\citenamefont {Ashida}\ \emph
  {et~al.}(2018{\natexlab{b}})\citenamefont {Ashida}, \citenamefont {Shi},
  \citenamefont {Ba\~nuls}, \citenamefont {Cirac},\ and\ \citenamefont
  {Demler}}]{PhysRevB.98.024103}%
  \BibitemOpen
  \bibfield  {author} {\bibinfo {author} {\bibfnamefont {Y.}~\bibnamefont
  {Ashida}}, \bibinfo {author} {\bibfnamefont {T.}~\bibnamefont {Shi}},
  \bibinfo {author} {\bibfnamefont {M.~C.}\ \bibnamefont {Ba\~nuls}}, \bibinfo
  {author} {\bibfnamefont {J.~I.}\ \bibnamefont {Cirac}},\ and\ \bibinfo
  {author} {\bibfnamefont {E.}~\bibnamefont {Demler}},\ }\bibfield  {title}
  {\bibinfo {title} {Variational principle for quantum impurity systems in and
  out of equilibrium: Application to kondo problems},\ }\href
  {https://doi.org/10.1103/PhysRevB.98.024103} {\bibfield  {journal} {\bibinfo
  {journal} {Phys. Rev. B}\ }\textbf {\bibinfo {volume} {98}},\ \bibinfo
  {pages} {024103} (\bibinfo {year} {2018}{\natexlab{b}})}\BibitemShut
  {NoStop}%
\bibitem [{\citenamefont {Park}\ \emph {et~al.}(2024)\citenamefont {Park},
  \citenamefont {Ng}, \citenamefont {Reichman},\ and\ \citenamefont
  {Chan}}]{Park_2024}%
  \BibitemOpen
  \bibfield  {author} {\bibinfo {author} {\bibfnamefont {G.}~\bibnamefont
  {Park}}, \bibinfo {author} {\bibfnamefont {N.}~\bibnamefont {Ng}}, \bibinfo
  {author} {\bibfnamefont {D.~R.}\ \bibnamefont {Reichman}},\ and\ \bibinfo
  {author} {\bibfnamefont {G.~K.-L.}\ \bibnamefont {Chan}},\ }\bibfield
  {title} {\bibinfo {title} {Tensor network influence functionals in the
  continuous-time limit: Connections to quantum embedding, bath discretization,
  and higher-order time propagation},\ }\bibfield  {journal} {\bibinfo
  {journal} {Physical Review B}\ }\textbf {\bibinfo {volume} {110}},\ \href
  {https://doi.org/10.1103/physrevb.110.045104} {10.1103/physrevb.110.045104}
  (\bibinfo {year} {2024})\BibitemShut {NoStop}%
\bibitem [{\citenamefont {Thoenniss}\ \emph {et~al.}(2025)\citenamefont
  {Thoenniss}, \citenamefont {Vilkoviskiy},\ and\ \citenamefont
  {Abanin}}]{Thoenniss_2025}%
  \BibitemOpen
  \bibfield  {author} {\bibinfo {author} {\bibfnamefont {J.}~\bibnamefont
  {Thoenniss}}, \bibinfo {author} {\bibfnamefont {I.}~\bibnamefont
  {Vilkoviskiy}},\ and\ \bibinfo {author} {\bibfnamefont {D.~A.}\ \bibnamefont
  {Abanin}},\ }\bibfield  {title} {\bibinfo {title} {Efficient pseudomode
  representation and complexity of quantum impurity models},\ }\bibfield
  {journal} {\bibinfo  {journal} {Physical Review B}\ }\textbf {\bibinfo
  {volume} {112}},\ \href {https://doi.org/10.1103/h8g7-bmng}
  {10.1103/h8g7-bmng} (\bibinfo {year} {2025})\BibitemShut {NoStop}%
\bibitem [{\citenamefont {Zgid}\ and\ \citenamefont
  {Chan}(2011)}]{Zgid_Chan_2011}%
  \BibitemOpen
  \bibfield  {author} {\bibinfo {author} {\bibfnamefont {D.}~\bibnamefont
  {Zgid}}\ and\ \bibinfo {author} {\bibfnamefont {G.~K.-L.}\ \bibnamefont
  {Chan}},\ }\bibfield  {title} {\bibinfo {title} {Dynamical mean-field theory
  from a quantum chemical perspective},\ }\href
  {https://doi.org/10.1063/1.3556707} {\bibfield  {journal} {\bibinfo
  {journal} {The Journal of Chemical Physics}\ }\textbf {\bibinfo {volume}
  {134}},\ \bibinfo {pages} {094115} (\bibinfo {year} {2011})}\BibitemShut
  {NoStop}%
\bibitem [{\citenamefont {He}\ and\ \citenamefont {Lu}(2014)}]{He_Lu_2014}%
  \BibitemOpen
  \bibfield  {author} {\bibinfo {author} {\bibfnamefont {R.-Q.}\ \bibnamefont
  {He}}\ and\ \bibinfo {author} {\bibfnamefont {Z.-Y.}\ \bibnamefont {Lu}},\
  }\bibfield  {title} {\bibinfo {title} {Quantum renormalization groups based
  on natural orbitals},\ }\href {https://doi.org/10.1103/PhysRevB.89.085108}
  {\bibfield  {journal} {\bibinfo  {journal} {Phys. Rev. B}\ }\textbf {\bibinfo
  {volume} {89}},\ \bibinfo {pages} {085108} (\bibinfo {year}
  {2014})}\BibitemShut {NoStop}%
\bibitem [{\citenamefont {Lu}\ \emph {et~al.}(2014)\citenamefont {Lu},
  \citenamefont {H\"oppner}, \citenamefont {Gunnarsson},\ and\ \citenamefont
  {Haverkort}}]{Lu_Hoppner_Gunnarsson_Haverkort_2014}%
  \BibitemOpen
  \bibfield  {author} {\bibinfo {author} {\bibfnamefont {Y.}~\bibnamefont
  {Lu}}, \bibinfo {author} {\bibfnamefont {M.}~\bibnamefont {H\"oppner}},
  \bibinfo {author} {\bibfnamefont {O.}~\bibnamefont {Gunnarsson}},\ and\
  \bibinfo {author} {\bibfnamefont {M.~W.}\ \bibnamefont {Haverkort}},\
  }\bibfield  {title} {\bibinfo {title} {Efficient real-frequency solver for
  dynamical mean-field theory},\ }\href
  {https://doi.org/10.1103/PhysRevB.90.085102} {\bibfield  {journal} {\bibinfo
  {journal} {Phys. Rev. B}\ }\textbf {\bibinfo {volume} {90}},\ \bibinfo
  {pages} {085102} (\bibinfo {year} {2014})}\BibitemShut {NoStop}%
\bibitem [{\citenamefont {Bravyi}\ and\ \citenamefont
  {Gosset}(2017)}]{Bravyi_Gosset_2017}%
  \BibitemOpen
  \bibfield  {author} {\bibinfo {author} {\bibfnamefont {S.}~\bibnamefont
  {Bravyi}}\ and\ \bibinfo {author} {\bibfnamefont {D.}~\bibnamefont
  {Gosset}},\ }\bibfield  {title} {\bibinfo {title} {Complexity of quantum
  impurity problems},\ }\href {https://doi.org/10.1007/s00220-017-2976-9}
  {\bibfield  {journal} {\bibinfo  {journal} {Communications in Mathematical
  Physics}\ }\textbf {\bibinfo {volume} {356}},\ \bibinfo {pages} {451–500}
  (\bibinfo {year} {2017})}\BibitemShut {NoStop}%
\bibitem [{\citenamefont {Boutin}\ and\ \citenamefont
  {Bauer}(2021)}]{Boutin_Bauer_2021}%
  \BibitemOpen
  \bibfield  {author} {\bibinfo {author} {\bibfnamefont {S.}~\bibnamefont
  {Boutin}}\ and\ \bibinfo {author} {\bibfnamefont {B.}~\bibnamefont {Bauer}},\
  }\bibfield  {title} {\bibinfo {title} {Quantum impurity models using
  superpositions of fermionic gaussian states: Practical methods and
  applications},\ }\href {https://doi.org/10.1103/PhysRevResearch.3.033188}
  {\bibfield  {journal} {\bibinfo  {journal} {Phys. Rev. Res.}\ }\textbf
  {\bibinfo {volume} {3}},\ \bibinfo {pages} {033188} (\bibinfo {year}
  {2021})}\BibitemShut {NoStop}%
\bibitem [{\citenamefont {Debertolis}\ \emph {et~al.}(2021)\citenamefont
  {Debertolis}, \citenamefont {Florens},\ and\ \citenamefont
  {Snyman}}]{Debertolis_Florens_Snyman_2021}%
  \BibitemOpen
  \bibfield  {author} {\bibinfo {author} {\bibfnamefont {M.}~\bibnamefont
  {Debertolis}}, \bibinfo {author} {\bibfnamefont {S.}~\bibnamefont
  {Florens}},\ and\ \bibinfo {author} {\bibfnamefont {I.}~\bibnamefont
  {Snyman}},\ }\bibfield  {title} {\bibinfo {title} {Few-body nature of kondo
  correlated ground states},\ }\href
  {https://doi.org/10.1103/PhysRevB.103.235166} {\bibfield  {journal} {\bibinfo
   {journal} {Phys. Rev. B}\ }\textbf {\bibinfo {volume} {103}},\ \bibinfo
  {pages} {235166} (\bibinfo {year} {2021})}\BibitemShut {NoStop}%
\bibitem [{\citenamefont {Debertolis}\ \emph {et~al.}(2022)\citenamefont
  {Debertolis}, \citenamefont {Snyman},\ and\ \citenamefont
  {Florens}}]{Debertolis_Snyman_Florens_2022}%
  \BibitemOpen
  \bibfield  {author} {\bibinfo {author} {\bibfnamefont {M.}~\bibnamefont
  {Debertolis}}, \bibinfo {author} {\bibfnamefont {I.}~\bibnamefont {Snyman}},\
  and\ \bibinfo {author} {\bibfnamefont {S.}~\bibnamefont {Florens}},\
  }\bibfield  {title} {\bibinfo {title} {Simulating realistic screening clouds
  around quantum impurities: Role of spatial anisotropy and disorder},\ }\href
  {https://doi.org/10.1103/PhysRevB.106.125115} {\bibfield  {journal} {\bibinfo
   {journal} {Phys. Rev. B}\ }\textbf {\bibinfo {volume} {106}},\ \bibinfo
  {pages} {125115} (\bibinfo {year} {2022})}\BibitemShut {NoStop}%
\bibitem [{\citenamefont {Bera}\ \emph {et~al.}(2015)\citenamefont {Bera},
  \citenamefont {Schomerus}, \citenamefont {Heidrich-Meisner},\ and\
  \citenamefont {Bardarson}}]{Bera_Schomerus_Meisner_Bardarson_2015}%
  \BibitemOpen
  \bibfield  {author} {\bibinfo {author} {\bibfnamefont {S.}~\bibnamefont
  {Bera}}, \bibinfo {author} {\bibfnamefont {H.}~\bibnamefont {Schomerus}},
  \bibinfo {author} {\bibfnamefont {F.}~\bibnamefont {Heidrich-Meisner}},\ and\
  \bibinfo {author} {\bibfnamefont {J.~H.}\ \bibnamefont {Bardarson}},\
  }\bibfield  {title} {\bibinfo {title} {Many-body localization characterized
  from a one-particle perspective},\ }\href
  {https://doi.org/10.1103/PhysRevLett.115.046603} {\bibfield  {journal}
  {\bibinfo  {journal} {Phys. Rev. Lett.}\ }\textbf {\bibinfo {volume} {115}},\
  \bibinfo {pages} {046603} (\bibinfo {year} {2015})}\BibitemShut {NoStop}%
\bibitem [{\citenamefont {Bera}\ \emph {et~al.}(2017)\citenamefont {Bera},
  \citenamefont {Martynec}, \citenamefont {Schomerus}, \citenamefont
  {Heidrich-Meisner},\ and\ \citenamefont
  {Bardarson}}]{Bera_Martynec_Schomerus_Meisner_Bardarson_2017}%
  \BibitemOpen
  \bibfield  {author} {\bibinfo {author} {\bibfnamefont {S.}~\bibnamefont
  {Bera}}, \bibinfo {author} {\bibfnamefont {T.}~\bibnamefont {Martynec}},
  \bibinfo {author} {\bibfnamefont {H.}~\bibnamefont {Schomerus}}, \bibinfo
  {author} {\bibfnamefont {F.}~\bibnamefont {Heidrich-Meisner}},\ and\ \bibinfo
  {author} {\bibfnamefont {J.~H.}\ \bibnamefont {Bardarson}},\ }\bibfield
  {title} {\bibinfo {title} {One-particle density matrix characterization of
  many-body localization},\ }\href
  {https://doi.org/https://doi.org/10.1002/andp.201600356} {\bibfield
  {journal} {\bibinfo  {journal} {Annalen der Physik}\ }\textbf {\bibinfo
  {volume} {529}},\ \bibinfo {pages} {1600356} (\bibinfo {year}
  {2017})}\BibitemShut {NoStop}%
\bibitem [{\citenamefont {Buijsman}\ \emph {et~al.}(2018)\citenamefont
  {Buijsman}, \citenamefont {Gritsev},\ and\ \citenamefont
  {Cheianov}}]{Buijsman_Gritsev_Cheianov_2018}%
  \BibitemOpen
  \bibfield  {author} {\bibinfo {author} {\bibfnamefont {W.}~\bibnamefont
  {Buijsman}}, \bibinfo {author} {\bibfnamefont {V.}~\bibnamefont {Gritsev}},\
  and\ \bibinfo {author} {\bibfnamefont {V.}~\bibnamefont {Cheianov}},\
  }\bibfield  {title} {\bibinfo {title} {Many-body localization in the fock
  space of natural orbitals},\ }\href
  {https://doi.org/10.21468/SciPostPhys.4.6.038} {\bibfield  {journal}
  {\bibinfo  {journal} {SciPost Phys.}\ }\textbf {\bibinfo {volume} {4}},\
  \bibinfo {pages} {038} (\bibinfo {year} {2018})}\BibitemShut {NoStop}%
\bibitem [{\citenamefont {Hopjan}\ and\ \citenamefont
  {Heidrich-Meisner}(2020)}]{Hopjan_Meisner_2020}%
  \BibitemOpen
  \bibfield  {author} {\bibinfo {author} {\bibfnamefont {M.}~\bibnamefont
  {Hopjan}}\ and\ \bibinfo {author} {\bibfnamefont {F.}~\bibnamefont
  {Heidrich-Meisner}},\ }\bibfield  {title} {\bibinfo {title} {Many-body
  localization from a one-particle perspective in the disordered
  one-dimensional bose-hubbard model},\ }\href
  {https://doi.org/10.1103/PhysRevA.101.063617} {\bibfield  {journal} {\bibinfo
   {journal} {Phys. Rev. A}\ }\textbf {\bibinfo {volume} {101}},\ \bibinfo
  {pages} {063617} (\bibinfo {year} {2020})}\BibitemShut {NoStop}%
\bibitem [{\citenamefont {Chitambar}\ and\ \citenamefont
  {Gour}(2019)}]{Chitambar_Gour_2019}%
  \BibitemOpen
  \bibfield  {author} {\bibinfo {author} {\bibfnamefont {E.}~\bibnamefont
  {Chitambar}}\ and\ \bibinfo {author} {\bibfnamefont {G.}~\bibnamefont
  {Gour}},\ }\bibfield  {title} {\bibinfo {title} {Quantum resource theories},\
  }\href {https://doi.org/10.1103/RevModPhys.91.025001} {\bibfield  {journal}
  {\bibinfo  {journal} {Rev. Mod. Phys.}\ }\textbf {\bibinfo {volume} {91}},\
  \bibinfo {pages} {025001} (\bibinfo {year} {2019})}\BibitemShut {NoStop}%
\bibitem [{\citenamefont {Gottesman}(1998)}]{Gottesman_1998}%
  \BibitemOpen
  \bibfield  {author} {\bibinfo {author} {\bibfnamefont {D.}~\bibnamefont
  {Gottesman}},\ }\bibfield  {title} {\bibinfo {title} {The heisenberg
  representation of quantum computers},\ }\href
  {https://arxiv.org/abs/quant-ph/9807006} {\bibfield  {journal} {\bibinfo
  {journal} {arXiv:quant-ph/9807006}\ } (\bibinfo {year} {1998})}\BibitemShut
  {NoStop}%
\bibitem [{\citenamefont {Aaronson}\ and\ \citenamefont
  {Gottesman}(2004)}]{Gottesmman_Aaronson_2004}%
  \BibitemOpen
  \bibfield  {author} {\bibinfo {author} {\bibfnamefont {S.}~\bibnamefont
  {Aaronson}}\ and\ \bibinfo {author} {\bibfnamefont {D.}~\bibnamefont
  {Gottesman}},\ }\bibfield  {title} {\bibinfo {title} {Improved simulation of
  stabilizer circuits},\ }\href {https://doi.org/10.1103/PhysRevA.70.052328}
  {\bibfield  {journal} {\bibinfo  {journal} {Phys. Rev. A}\ }\textbf {\bibinfo
  {volume} {70}},\ \bibinfo {pages} {052328} (\bibinfo {year}
  {2004})}\BibitemShut {NoStop}%
\bibitem [{\citenamefont {Barenco}\ \emph {et~al.}(1995)\citenamefont
  {Barenco}, \citenamefont {Bennett}, \citenamefont {Cleve}, \citenamefont
  {DiVincenzo}, \citenamefont {Margolus}, \citenamefont {Shor}, \citenamefont
  {Sleator}, \citenamefont {Smolin},\ and\ \citenamefont
  {Weinfurter}}]{Barenco_Bennett_etal_1995}%
  \BibitemOpen
  \bibfield  {author} {\bibinfo {author} {\bibfnamefont {A.}~\bibnamefont
  {Barenco}}, \bibinfo {author} {\bibfnamefont {C.~H.}\ \bibnamefont
  {Bennett}}, \bibinfo {author} {\bibfnamefont {R.}~\bibnamefont {Cleve}},
  \bibinfo {author} {\bibfnamefont {D.~P.}\ \bibnamefont {DiVincenzo}},
  \bibinfo {author} {\bibfnamefont {N.}~\bibnamefont {Margolus}}, \bibinfo
  {author} {\bibfnamefont {P.}~\bibnamefont {Shor}}, \bibinfo {author}
  {\bibfnamefont {T.}~\bibnamefont {Sleator}}, \bibinfo {author} {\bibfnamefont
  {J.~A.}\ \bibnamefont {Smolin}},\ and\ \bibinfo {author} {\bibfnamefont
  {H.}~\bibnamefont {Weinfurter}},\ }\bibfield  {title} {\bibinfo {title}
  {Elementary gates for quantum computation},\ }\href
  {https://doi.org/10.1103/PhysRevA.52.3457} {\bibfield  {journal} {\bibinfo
  {journal} {Phys. Rev. A}\ }\textbf {\bibinfo {volume} {52}},\ \bibinfo
  {pages} {3457} (\bibinfo {year} {1995})}\BibitemShut {NoStop}%
\bibitem [{\citenamefont {Bravyi}\ and\ \citenamefont
  {Kitaev}(2005)}]{Bravyi_Kitaev_2005}%
  \BibitemOpen
  \bibfield  {author} {\bibinfo {author} {\bibfnamefont {S.}~\bibnamefont
  {Bravyi}}\ and\ \bibinfo {author} {\bibfnamefont {A.}~\bibnamefont
  {Kitaev}},\ }\bibfield  {title} {\bibinfo {title} {Universal quantum
  computation with ideal clifford gates and noisy ancillas},\ }\href
  {https://doi.org/10.1103/PhysRevA.71.022316} {\bibfield  {journal} {\bibinfo
  {journal} {Phys. Rev. A}\ }\textbf {\bibinfo {volume} {71}},\ \bibinfo
  {pages} {022316} (\bibinfo {year} {2005})}\BibitemShut {NoStop}%
\bibitem [{\citenamefont {Veitch}\ \emph {et~al.}(2014)\citenamefont {Veitch},
  \citenamefont {Hamed~Mousavian}, \citenamefont {Gottesman},\ and\
  \citenamefont {Emerson}}]{Veitch_2014}%
  \BibitemOpen
  \bibfield  {author} {\bibinfo {author} {\bibfnamefont {V.}~\bibnamefont
  {Veitch}}, \bibinfo {author} {\bibfnamefont {S.~A.}\ \bibnamefont
  {Hamed~Mousavian}}, \bibinfo {author} {\bibfnamefont {D.}~\bibnamefont
  {Gottesman}},\ and\ \bibinfo {author} {\bibfnamefont {J.}~\bibnamefont
  {Emerson}},\ }\bibfield  {title} {\bibinfo {title} {The resource theory of
  stabilizer quantum computation},\ }\href
  {https://doi.org/10.1088/1367-2630/16/1/013009} {\bibfield  {journal}
  {\bibinfo  {journal} {New Journal of Physics}\ }\textbf {\bibinfo {volume}
  {16}},\ \bibinfo {pages} {013009} (\bibinfo {year} {2014})}\BibitemShut
  {NoStop}%
\bibitem [{\citenamefont {Bravyi}\ and\ \citenamefont
  {Gosset}(2016)}]{Bravyi_Gosset_2016}%
  \BibitemOpen
  \bibfield  {author} {\bibinfo {author} {\bibfnamefont {S.}~\bibnamefont
  {Bravyi}}\ and\ \bibinfo {author} {\bibfnamefont {D.}~\bibnamefont
  {Gosset}},\ }\bibfield  {title} {\bibinfo {title} {Improved classical
  simulation of quantum circuits dominated by clifford gates},\ }\href
  {https://doi.org/10.1103/PhysRevLett.116.250501} {\bibfield  {journal}
  {\bibinfo  {journal} {Phys. Rev. Lett.}\ }\textbf {\bibinfo {volume} {116}},\
  \bibinfo {pages} {250501} (\bibinfo {year} {2016})}\BibitemShut {NoStop}%
\bibitem [{\citenamefont {Bravyi}\ \emph {et~al.}(2016)\citenamefont {Bravyi},
  \citenamefont {Smith},\ and\ \citenamefont
  {Smolin}}]{Bravyi_Smith_Smolin_2016}%
  \BibitemOpen
  \bibfield  {author} {\bibinfo {author} {\bibfnamefont {S.}~\bibnamefont
  {Bravyi}}, \bibinfo {author} {\bibfnamefont {G.}~\bibnamefont {Smith}},\ and\
  \bibinfo {author} {\bibfnamefont {J.~A.}\ \bibnamefont {Smolin}},\ }\bibfield
   {title} {\bibinfo {title} {Trading classical and quantum computational
  resources},\ }\href {https://doi.org/10.1103/PhysRevX.6.021043} {\bibfield
  {journal} {\bibinfo  {journal} {Phys. Rev. X}\ }\textbf {\bibinfo {volume}
  {6}},\ \bibinfo {pages} {021043} (\bibinfo {year} {2016})}\BibitemShut
  {NoStop}%
\bibitem [{\citenamefont {Hamaguchi}\ \emph {et~al.}(2024)\citenamefont
  {Hamaguchi}, \citenamefont {Hamada},\ and\ \citenamefont
  {Yoshioka}}]{Hamaguchi2024handbookquantifying}%
  \BibitemOpen
  \bibfield  {author} {\bibinfo {author} {\bibfnamefont {H.}~\bibnamefont
  {Hamaguchi}}, \bibinfo {author} {\bibfnamefont {K.}~\bibnamefont {Hamada}},\
  and\ \bibinfo {author} {\bibfnamefont {N.}~\bibnamefont {Yoshioka}},\
  }\bibfield  {title} {\bibinfo {title} {Handbook for {Q}uantifying
  {R}obustness of {M}agic},\ }\href
  {https://doi.org/10.22331/q-2024-09-05-1461} {\bibfield  {journal} {\bibinfo
  {journal} {{Quantum}}\ }\textbf {\bibinfo {volume} {8}},\ \bibinfo {pages}
  {1461} (\bibinfo {year} {2024})}\BibitemShut {NoStop}%
\bibitem [{\citenamefont {Collura}\ \emph {et~al.}(2025)\citenamefont
  {Collura}, \citenamefont {Nardis}, \citenamefont {Alba},\ and\ \citenamefont
  {Lami}}]{Collura_DeNardis_Alba_Lami_2025}%
  \BibitemOpen
  \bibfield  {author} {\bibinfo {author} {\bibfnamefont {M.}~\bibnamefont
  {Collura}}, \bibinfo {author} {\bibfnamefont {J.~D.}\ \bibnamefont {Nardis}},
  \bibinfo {author} {\bibfnamefont {V.}~\bibnamefont {Alba}},\ and\ \bibinfo
  {author} {\bibfnamefont {G.}~\bibnamefont {Lami}},\ }\bibfield  {title}
  {\bibinfo {title} {The quantum magic of fermionic gaussian states},\ }\href
  {https://arxiv.org/abs/2412.05367} {\bibfield  {journal} {\bibinfo  {journal}
  {arXiv:2412.05367}\ } (\bibinfo {year} {2025})}\BibitemShut {NoStop}%
\bibitem [{\citenamefont {Sierant}\ \emph {et~al.}(2026)\citenamefont
  {Sierant}, \citenamefont {Stornati},\ and\ \citenamefont
  {Turkeshi}}]{Sierant_Stornati_Turkeshi_2025}%
  \BibitemOpen
  \bibfield  {author} {\bibinfo {author} {\bibfnamefont {P.}~\bibnamefont
  {Sierant}}, \bibinfo {author} {\bibfnamefont {P.}~\bibnamefont {Stornati}},\
  and\ \bibinfo {author} {\bibfnamefont {X.}~\bibnamefont {Turkeshi}},\
  }\bibfield  {title} {\bibinfo {title} {Fermionic magic resources of quantum
  many-body systems},\ }\href
  {https://doi.org/https://doi.org/10.1103/3yx4-1j27} {\bibfield  {journal}
  {\bibinfo  {journal} {PRX Quantum}\ }\textbf {\bibinfo {volume} {7}},\
  \bibinfo {pages} {010302} (\bibinfo {year} {2026})}\BibitemShut {NoStop}%
\bibitem [{\citenamefont {Leone}\ \emph {et~al.}(2022)\citenamefont {Leone},
  \citenamefont {Oliviero},\ and\ \citenamefont
  {Hamma}}]{Leone_Oliviero_Hamma_2022}%
  \BibitemOpen
  \bibfield  {author} {\bibinfo {author} {\bibfnamefont {L.}~\bibnamefont
  {Leone}}, \bibinfo {author} {\bibfnamefont {S.~F.~E.}\ \bibnamefont
  {Oliviero}},\ and\ \bibinfo {author} {\bibfnamefont {A.}~\bibnamefont
  {Hamma}},\ }\bibfield  {title} {\bibinfo {title} {Stabilizer r\'enyi
  entropy},\ }\href {https://doi.org/10.1103/PhysRevLett.128.050402} {\bibfield
   {journal} {\bibinfo  {journal} {Phys. Rev. Lett.}\ }\textbf {\bibinfo
  {volume} {128}},\ \bibinfo {pages} {050402} (\bibinfo {year}
  {2022})}\BibitemShut {NoStop}%
\bibitem [{\citenamefont {Haug}\ and\ \citenamefont
  {Piroli}(2023)}]{Haug_Piroli_2023}%
  \BibitemOpen
  \bibfield  {author} {\bibinfo {author} {\bibfnamefont {T.}~\bibnamefont
  {Haug}}\ and\ \bibinfo {author} {\bibfnamefont {L.}~\bibnamefont {Piroli}},\
  }\bibfield  {title} {\bibinfo {title} {Quantifying nonstabilizerness of
  matrix product states},\ }\href {https://doi.org/10.1103/PhysRevB.107.035148}
  {\bibfield  {journal} {\bibinfo  {journal} {Phys. Rev. B}\ }\textbf {\bibinfo
  {volume} {107}},\ \bibinfo {pages} {035148} (\bibinfo {year}
  {2023})}\BibitemShut {NoStop}%
\bibitem [{\citenamefont {Lami}\ and\ \citenamefont
  {Collura}(2023)}]{Lami_Collura_2023}%
  \BibitemOpen
  \bibfield  {author} {\bibinfo {author} {\bibfnamefont {G.}~\bibnamefont
  {Lami}}\ and\ \bibinfo {author} {\bibfnamefont {M.}~\bibnamefont {Collura}},\
  }\bibfield  {title} {\bibinfo {title} {Nonstabilizerness via perfect pauli
  sampling of matrix product states},\ }\href
  {https://doi.org/10.1103/PhysRevLett.131.180401} {\bibfield  {journal}
  {\bibinfo  {journal} {Phys. Rev. Lett.}\ }\textbf {\bibinfo {volume} {131}},\
  \bibinfo {pages} {180401} (\bibinfo {year} {2023})}\BibitemShut {NoStop}%
\bibitem [{\citenamefont {Leone}\ and\ \citenamefont
  {Bittel}(2024)}]{Leone_Bittel_2024}%
  \BibitemOpen
  \bibfield  {author} {\bibinfo {author} {\bibfnamefont {L.}~\bibnamefont
  {Leone}}\ and\ \bibinfo {author} {\bibfnamefont {L.}~\bibnamefont {Bittel}},\
  }\bibfield  {title} {\bibinfo {title} {Stabilizer entropies are monotones for
  magic-state resource theory},\ }\href
  {https://doi.org/10.1103/PhysRevA.110.L040403} {\bibfield  {journal}
  {\bibinfo  {journal} {Phys. Rev. A}\ }\textbf {\bibinfo {volume} {110}},\
  \bibinfo {pages} {L040403} (\bibinfo {year} {2024})}\BibitemShut {NoStop}%
\bibitem [{\citenamefont {Tarabunga}\ \emph {et~al.}(2024)\citenamefont
  {Tarabunga}, \citenamefont {Tirrito}, \citenamefont {Ba\~nuls},\ and\
  \citenamefont {Dalmonte}}]{Tarabunga_Tirrito_Banuls_Dalmonte_2024}%
  \BibitemOpen
  \bibfield  {author} {\bibinfo {author} {\bibfnamefont {P.~S.}\ \bibnamefont
  {Tarabunga}}, \bibinfo {author} {\bibfnamefont {E.}~\bibnamefont {Tirrito}},
  \bibinfo {author} {\bibfnamefont {M.~C.}\ \bibnamefont {Ba\~nuls}},\ and\
  \bibinfo {author} {\bibfnamefont {M.}~\bibnamefont {Dalmonte}},\ }\bibfield
  {title} {\bibinfo {title} {Nonstabilizerness via matrix product states in the
  pauli basis},\ }\href {https://doi.org/10.1103/PhysRevLett.133.010601}
  {\bibfield  {journal} {\bibinfo  {journal} {Phys. Rev. Lett.}\ }\textbf
  {\bibinfo {volume} {133}},\ \bibinfo {pages} {010601} (\bibinfo {year}
  {2024})}\BibitemShut {NoStop}%
\bibitem [{\citenamefont {Terhal}\ and\ \citenamefont
  {DiVincenzo}(2002)}]{Terhal_DiVincenzo_2002}%
  \BibitemOpen
  \bibfield  {author} {\bibinfo {author} {\bibfnamefont {B.~M.}\ \bibnamefont
  {Terhal}}\ and\ \bibinfo {author} {\bibfnamefont {D.~P.}\ \bibnamefont
  {DiVincenzo}},\ }\bibfield  {title} {\bibinfo {title} {Classical simulation
  of noninteracting-fermion quantum circuits},\ }\href
  {https://doi.org/10.1103/PhysRevA.65.032325} {\bibfield  {journal} {\bibinfo
  {journal} {Phys. Rev. A}\ }\textbf {\bibinfo {volume} {65}},\ \bibinfo
  {pages} {032325} (\bibinfo {year} {2002})}\BibitemShut {NoStop}%
\bibitem [{\citenamefont {Bravyi}(2004)}]{Bravy_2004_Lagrangian}%
  \BibitemOpen
  \bibfield  {author} {\bibinfo {author} {\bibfnamefont {S.}~\bibnamefont
  {Bravyi}},\ }\bibfield  {title} {\bibinfo {title} {Lagrangian representation
  for fermionic linear optics},\ }\href
  {https://arxiv.org/abs/quant-ph/0404180} {\bibfield  {journal} {\bibinfo
  {journal} {arXiv:quant-ph/0404180}\ } (\bibinfo {year} {2004})}\BibitemShut
  {NoStop}%
\bibitem [{\citenamefont {Reardon-Smith}\ \emph {et~al.}(2024)\citenamefont
  {Reardon-Smith}, \citenamefont {Oszmaniec},\ and\ \citenamefont
  {Korzekwa}}]{ReardonSmith_Oszmaniec_Korzekwa_2024}%
  \BibitemOpen
  \bibfield  {author} {\bibinfo {author} {\bibfnamefont {O.}~\bibnamefont
  {Reardon-Smith}}, \bibinfo {author} {\bibfnamefont {M.}~\bibnamefont
  {Oszmaniec}},\ and\ \bibinfo {author} {\bibfnamefont {K.}~\bibnamefont
  {Korzekwa}},\ }\bibfield  {title} {\bibinfo {title} {Improved simulation of
  quantum circuits dominated by free fermionic operations},\ }\href
  {https://doi.org/10.22331/q-2024-12-04-1549} {\bibfield  {journal} {\bibinfo
  {journal} {{Quantum}}\ }\textbf {\bibinfo {volume} {8}},\ \bibinfo {pages}
  {1549} (\bibinfo {year} {2024})}\BibitemShut {NoStop}%
\bibitem [{\citenamefont {Dias}\ and\ \citenamefont
  {Koenig}(2024)}]{Dias_Koenig_2024}%
  \BibitemOpen
  \bibfield  {author} {\bibinfo {author} {\bibfnamefont {B.}~\bibnamefont
  {Dias}}\ and\ \bibinfo {author} {\bibfnamefont {R.}~\bibnamefont {Koenig}},\
  }\bibfield  {title} {\bibinfo {title} {Classical simulation of non-{G}aussian
  fermionic circuits},\ }\href {https://doi.org/10.22331/q-2024-05-21-1350}
  {\bibfield  {journal} {\bibinfo  {journal} {{Quantum}}\ }\textbf {\bibinfo
  {volume} {8}},\ \bibinfo {pages} {1350} (\bibinfo {year} {2024})}\BibitemShut
  {NoStop}%
\bibitem [{\citenamefont {Lyu}\ and\ \citenamefont {Bu}(2024)}]{Lyu_Bu_2024}%
  \BibitemOpen
  \bibfield  {author} {\bibinfo {author} {\bibfnamefont {X.}~\bibnamefont
  {Lyu}}\ and\ \bibinfo {author} {\bibfnamefont {K.}~\bibnamefont {Bu}},\
  }\bibfield  {title} {\bibinfo {title} {Fermionic gaussian testing and
  non-gaussian measures via convolution},\ }\href
  {https://arxiv.org/abs/2409.08180} {\bibfield  {journal} {\bibinfo  {journal}
  {arXiv:2409.08180}\ } (\bibinfo {year} {2024})}\BibitemShut {NoStop}%
\bibitem [{\citenamefont {Coffman}\ \emph {et~al.}(2025)\citenamefont
  {Coffman}, \citenamefont {Smith},\ and\ \citenamefont
  {Gao}}]{Coffman_Smith_Gao_2025}%
  \BibitemOpen
  \bibfield  {author} {\bibinfo {author} {\bibfnamefont {L.}~\bibnamefont
  {Coffman}}, \bibinfo {author} {\bibfnamefont {G.}~\bibnamefont {Smith}},\
  and\ \bibinfo {author} {\bibfnamefont {X.}~\bibnamefont {Gao}},\ }\bibfield
  {title} {\bibinfo {title} {Measuring non-gaussian magic in fermions:
  Convolution, entropy, and the violation of wick's theorem and the matchgate
  identity},\ }\href {https://arxiv.org/abs/2501.06179} {\bibfield  {journal}
  {\bibinfo  {journal} {arXiv:2501.06179}\ } (\bibinfo {year}
  {2025})}\BibitemShut {NoStop}%
\bibitem [{\citenamefont {Dowling}\ \emph {et~al.}(2025)\citenamefont
  {Dowling}, \citenamefont {Kos},\ and\ \citenamefont
  {Turkeshi}}]{Dowling_Kos_Turkeshi_2024}%
  \BibitemOpen
  \bibfield  {author} {\bibinfo {author} {\bibfnamefont {N.}~\bibnamefont
  {Dowling}}, \bibinfo {author} {\bibfnamefont {P.}~\bibnamefont {Kos}},\ and\
  \bibinfo {author} {\bibfnamefont {X.}~\bibnamefont {Turkeshi}},\ }\bibfield
  {title} {\bibinfo {title} {Magic of the heisenberg picture},\ }\href
  {https://arxiv.org/abs/2408.16047} {\bibfield  {journal} {\bibinfo  {journal}
  {arXiv:2408.16047}\ } (\bibinfo {year} {2025})}\BibitemShut {NoStop}%
\bibitem [{\citenamefont {L\"owdin}(1955)}]{Lowdin_1955}%
  \BibitemOpen
  \bibfield  {author} {\bibinfo {author} {\bibfnamefont {P.-O.}\ \bibnamefont
  {L\"owdin}},\ }\bibfield  {title} {\bibinfo {title} {Quantum theory of
  many-particle systems. i. physical interpretations by means of density
  matrices, natural spin-orbitals, and convergence problems in the method of
  configurational interaction},\ }\href
  {https://doi.org/10.1103/PhysRev.97.1474} {\bibfield  {journal} {\bibinfo
  {journal} {Phys. Rev.}\ }\textbf {\bibinfo {volume} {97}},\ \bibinfo {pages}
  {1474} (\bibinfo {year} {1955})}\BibitemShut {NoStop}%
\bibitem [{\citenamefont {Davidson}(1972)}]{Davidson_1972}%
  \BibitemOpen
  \bibfield  {author} {\bibinfo {author} {\bibfnamefont {E.~R.}\ \bibnamefont
  {Davidson}},\ }\bibfield  {title} {\bibinfo {title} {Natural orbitals},\
  }\href {https://doi.org/https://doi.org/10.1016/S0065-3276(08)60547-X}
  {\bibfield  {journal} {\bibinfo  {journal} {Advances in Quantum Chemistry}\
  }\textbf {\bibinfo {volume} {6}},\ \bibinfo {pages} {235} (\bibinfo {year}
  {1972})}\BibitemShut {NoStop}%
\bibitem [{\citenamefont {Roos}\ \emph {et~al.}(1980)\citenamefont {Roos},
  \citenamefont {Taylor},\ and\ \citenamefont
  {Sigbahn}}]{Roos_Taylor_Sigbahn_1980}%
  \BibitemOpen
  \bibfield  {author} {\bibinfo {author} {\bibfnamefont {B.~O.}\ \bibnamefont
  {Roos}}, \bibinfo {author} {\bibfnamefont {P.~R.}\ \bibnamefont {Taylor}},\
  and\ \bibinfo {author} {\bibfnamefont {P.~E.}\ \bibnamefont {Sigbahn}},\
  }\bibfield  {title} {\bibinfo {title} {A complete active space scf method
  (casscf) using a density matrix formulated super-ci approach},\ }\href
  {https://doi.org/https://doi.org/10.1016/0301-0104(80)80045-0} {\bibfield
  {journal} {\bibinfo  {journal} {Chemical Physics}\ }\textbf {\bibinfo
  {volume} {48}},\ \bibinfo {pages} {157} (\bibinfo {year} {1980})}\BibitemShut
  {NoStop}%
\bibitem [{\citenamefont {Prosen}(2008)}]{prosen2008}%
  \BibitemOpen
  \bibfield  {author} {\bibinfo {author} {\bibfnamefont {T.}~\bibnamefont
  {Prosen}},\ }\bibfield  {title} {\bibinfo {title} {Third quantization: a
  general method to solve master equations for quadratic open {F}ermi
  systems},\ }\href {https://doi.org/10.1088/1367-2630/10/4/043026} {\bibfield
  {journal} {\bibinfo  {journal} {New J. Phys.}\ }\textbf {\bibinfo {volume}
  {10}},\ \bibinfo {pages} {043026} (\bibinfo {year} {2008})}\BibitemShut
  {NoStop}%
\bibitem [{\citenamefont {Penrose}\ and\ \citenamefont
  {Onsager}(1956)}]{PhysRev.104.576}%
  \BibitemOpen
  \bibfield  {author} {\bibinfo {author} {\bibfnamefont {O.}~\bibnamefont
  {Penrose}}\ and\ \bibinfo {author} {\bibfnamefont {L.}~\bibnamefont
  {Onsager}},\ }\bibfield  {title} {\bibinfo {title} {Bose-einstein
  condensation and liquid helium},\ }\href
  {https://doi.org/10.1103/PhysRev.104.576} {\bibfield  {journal} {\bibinfo
  {journal} {Phys. Rev.}\ }\textbf {\bibinfo {volume} {104}},\ \bibinfo {pages}
  {576} (\bibinfo {year} {1956})}\BibitemShut {NoStop}%
\bibitem [{\citenamefont {Benavides-Riveros}\ \emph {et~al.}(2020)\citenamefont
  {Benavides-Riveros}, \citenamefont {Wolff}, \citenamefont {Marques},\ and\
  \citenamefont {Schilling}}]{PhysRevLett.124.180603}%
  \BibitemOpen
  \bibfield  {author} {\bibinfo {author} {\bibfnamefont {C.~L.}\ \bibnamefont
  {Benavides-Riveros}}, \bibinfo {author} {\bibfnamefont {J.}~\bibnamefont
  {Wolff}}, \bibinfo {author} {\bibfnamefont {M.~A.~L.}\ \bibnamefont
  {Marques}},\ and\ \bibinfo {author} {\bibfnamefont {C.}~\bibnamefont
  {Schilling}},\ }\bibfield  {title} {\bibinfo {title} {Reduced density matrix
  functional theory for bosons},\ }\href
  {https://doi.org/10.1103/PhysRevLett.124.180603} {\bibfield  {journal}
  {\bibinfo  {journal} {Phys. Rev. Lett.}\ }\textbf {\bibinfo {volume} {124}},\
  \bibinfo {pages} {180603} (\bibinfo {year} {2020})}\BibitemShut {NoStop}%
\bibitem [{\citenamefont {García-March}\ \emph {et~al.}(2016)\citenamefont
  {García-March}, \citenamefont {Dehkharghani},\ and\ \citenamefont
  {Zinner}}]{Garc_a_March_2016}%
  \BibitemOpen
  \bibfield  {author} {\bibinfo {author} {\bibfnamefont {M.~A.}\ \bibnamefont
  {García-March}}, \bibinfo {author} {\bibfnamefont {A.~S.}\ \bibnamefont
  {Dehkharghani}},\ and\ \bibinfo {author} {\bibfnamefont {N.~T.}\ \bibnamefont
  {Zinner}},\ }\bibfield  {title} {\bibinfo {title} {Entanglement of an
  impurity in a few-body one-dimensional ideal bose system},\ }\href
  {https://doi.org/10.1088/0953-4075/49/7/075303} {\bibfield  {journal}
  {\bibinfo  {journal} {Journal of Physics B: Atomic, Molecular and Optical
  Physics}\ }\textbf {\bibinfo {volume} {49}},\ \bibinfo {pages} {075303}
  (\bibinfo {year} {2016})}\BibitemShut {NoStop}%
\bibitem [{\citenamefont {Rath}\ \emph {et~al.}(2023)\citenamefont {Rath},
  \citenamefont {Vitale}, \citenamefont {Murciano}, \citenamefont {Votto},
  \citenamefont {Dubail}, \citenamefont {Kueng}, \citenamefont {Branciard},
  \citenamefont {Calabrese},\ and\ \citenamefont
  {Vermersch}}]{Rath_Vitale_Murciano_Votto_Dubail_Kueng_Calabrese_Vermersch_2023}%
  \BibitemOpen
  \bibfield  {author} {\bibinfo {author} {\bibfnamefont {A.}~\bibnamefont
  {Rath}}, \bibinfo {author} {\bibfnamefont {V.}~\bibnamefont {Vitale}},
  \bibinfo {author} {\bibfnamefont {S.}~\bibnamefont {Murciano}}, \bibinfo
  {author} {\bibfnamefont {M.}~\bibnamefont {Votto}}, \bibinfo {author}
  {\bibfnamefont {J.}~\bibnamefont {Dubail}}, \bibinfo {author} {\bibfnamefont
  {R.}~\bibnamefont {Kueng}}, \bibinfo {author} {\bibfnamefont
  {C.}~\bibnamefont {Branciard}}, \bibinfo {author} {\bibfnamefont
  {P.}~\bibnamefont {Calabrese}},\ and\ \bibinfo {author} {\bibfnamefont
  {B.}~\bibnamefont {Vermersch}},\ }\bibfield  {title} {\bibinfo {title}
  {Entanglement barrier and its symmetry resolution: Theory and experimental
  observation},\ }\href {https://doi.org/10.1103/PRXQuantum.4.010318}
  {\bibfield  {journal} {\bibinfo  {journal} {PRX Quantum}\ }\textbf {\bibinfo
  {volume} {4}},\ \bibinfo {pages} {010318} (\bibinfo {year}
  {2023})}\BibitemShut {NoStop}%
\bibitem [{\citenamefont {Siva}\ \emph {et~al.}(2022)\citenamefont {Siva},
  \citenamefont {Zou}, \citenamefont {Soejima}, \citenamefont {Mong},\ and\
  \citenamefont {Zaletel}}]{Siva_Zou_Soejima_Mong_Zaletel_2022}%
  \BibitemOpen
  \bibfield  {author} {\bibinfo {author} {\bibfnamefont {K.}~\bibnamefont
  {Siva}}, \bibinfo {author} {\bibfnamefont {Y.}~\bibnamefont {Zou}}, \bibinfo
  {author} {\bibfnamefont {T.}~\bibnamefont {Soejima}}, \bibinfo {author}
  {\bibfnamefont {R.~S.~K.}\ \bibnamefont {Mong}},\ and\ \bibinfo {author}
  {\bibfnamefont {M.~P.}\ \bibnamefont {Zaletel}},\ }\bibfield  {title}
  {\bibinfo {title} {Universal tripartite entanglement signature of ungappable
  edge states},\ }\href {https://doi.org/10.1103/PhysRevB.106.L041107}
  {\bibfield  {journal} {\bibinfo  {journal} {Phys. Rev. B}\ }\textbf {\bibinfo
  {volume} {106}},\ \bibinfo {pages} {L041107} (\bibinfo {year}
  {2022})}\BibitemShut {NoStop}%
\bibitem [{\citenamefont {Haug}\ \emph {et~al.}(2026)\citenamefont {Haug},
  \citenamefont {Turkeshi},\ and\ \citenamefont {Sierant}}]{haug2026}%
  \BibitemOpen
  \bibfield  {author} {\bibinfo {author} {\bibfnamefont {T.}~\bibnamefont
  {Haug}}, \bibinfo {author} {\bibfnamefont {X.}~\bibnamefont {Turkeshi}},\
  and\ \bibinfo {author} {\bibfnamefont {P.}~\bibnamefont {Sierant}},\
  }\bibfield  {title} {\bibinfo {title} {Practical tests and witnesses of
  fermionic non-{G}aussianity},\ }\href {https://arxiv.org/abs/2605.26218}
  {\bibfield  {journal} {\bibinfo  {journal} {arXiv:2605.26218}\ } (\bibinfo
  {year} {2026})}\BibitemShut {NoStop}%
\bibitem [{\citenamefont {Tarabunga}\ \emph {et~al.}(2026)\citenamefont
  {Tarabunga}, \citenamefont {Jobst}, \citenamefont {Morral-Yepes},
  \citenamefont {Langer}, \citenamefont {Kraus}, \citenamefont {Pollmann},\
  and\ \citenamefont {Lin}}]{tarabunga2026cov}%
  \BibitemOpen
  \bibfield  {author} {\bibinfo {author} {\bibfnamefont {P.~S.}\ \bibnamefont
  {Tarabunga}}, \bibinfo {author} {\bibfnamefont {B.}~\bibnamefont {Jobst}},
  \bibinfo {author} {\bibfnamefont {R.}~\bibnamefont {Morral-Yepes}}, \bibinfo
  {author} {\bibfnamefont {M.}~\bibnamefont {Langer}}, \bibinfo {author}
  {\bibfnamefont {B.}~\bibnamefont {Kraus}}, \bibinfo {author} {\bibfnamefont
  {F.}~\bibnamefont {Pollmann}},\ and\ \bibinfo {author} {\bibfnamefont
  {S.-H.}\ \bibnamefont {Lin}},\ }\bibfield  {title} {\bibinfo {title}
  {Computable measures of fermionic non-{G}aussianity from the covariance
  matrix},\ }\href {https://arxiv.org/abs/2607.02242} {\bibfield  {journal}
  {\bibinfo  {journal} {arXiv:2607.02242}\ } (\bibinfo {year}
  {2026})}\BibitemShut {NoStop}%
\bibitem [{\citenamefont {Thouless}(1960)}]{Thouless_1960}%
  \BibitemOpen
  \bibfield  {author} {\bibinfo {author} {\bibfnamefont {D.}~\bibnamefont
  {Thouless}},\ }\bibfield  {title} {\bibinfo {title} {Stability conditions and
  nuclear rotations in the hartree-fock theory},\ }\href
  {https://doi.org/https://doi.org/10.1016/0029-5582(60)90048-1} {\bibfield
  {journal} {\bibinfo  {journal} {Nuclear Physics}\ }\textbf {\bibinfo {volume}
  {21}},\ \bibinfo {pages} {225} (\bibinfo {year} {1960})}\BibitemShut
  {NoStop}%
\bibitem [{\citenamefont {Collins}\ and\ \citenamefont
  {Nechita}(2010)}]{Collins2010_Weingarten}%
  \BibitemOpen
  \bibfield  {author} {\bibinfo {author} {\bibfnamefont {B.}~\bibnamefont
  {Collins}}\ and\ \bibinfo {author} {\bibfnamefont {I.}~\bibnamefont
  {Nechita}},\ }\bibfield  {title} {\bibinfo {title} {{Random quantum channels
  I: Graphical calculus and the Bell state phenomenon}},\ }\href
  {https://doi.org/10.1007/s00220-010-1012-0} {\bibfield  {journal} {\bibinfo
  {journal} {Communications in Mathematical Physics}\ }\textbf {\bibinfo
  {volume} {297}},\ \bibinfo {pages} {345} (\bibinfo {year}
  {2010})}\BibitemShut {NoStop}%
\bibitem [{\citenamefont {Vidal}(2004)}]{Vidal_2004}%
  \BibitemOpen
  \bibfield  {author} {\bibinfo {author} {\bibfnamefont {G.}~\bibnamefont
  {Vidal}},\ }\bibfield  {title} {\bibinfo {title} {Efficient simulation of
  one-dimensional quantum many-body systems},\ }\href
  {https://doi.org/10.1103/PhysRevLett.93.040502} {\bibfield  {journal}
  {\bibinfo  {journal} {Phys. Rev. Lett.}\ }\textbf {\bibinfo {volume} {93}},\
  \bibinfo {pages} {040502} (\bibinfo {year} {2004})}\BibitemShut {NoStop}%
\bibitem [{\citenamefont {Jordan}\ and\ \citenamefont
  {Wigner}(1928)}]{Jordan_Wigner_1928}%
  \BibitemOpen
  \bibfield  {author} {\bibinfo {author} {\bibfnamefont {P.}~\bibnamefont
  {Jordan}}\ and\ \bibinfo {author} {\bibfnamefont {E.}~\bibnamefont
  {Wigner}},\ }\bibfield  {title} {\bibinfo {title} {{\"U}ber das paulische
  {\"a}quivalenzverbot},\ }\href {https://doi.org/10.1007/BF01331938}
  {\bibfield  {journal} {\bibinfo  {journal} {Zeitschrift f{\"u}r Physik}\
  }\textbf {\bibinfo {volume} {47}},\ \bibinfo {pages} {631} (\bibinfo {year}
  {1928})}\BibitemShut {NoStop}%
\bibitem [{\citenamefont {Khemani}\ \emph {et~al.}(2018)\citenamefont
  {Khemani}, \citenamefont {Huse},\ and\ \citenamefont
  {Nahum}}]{Khemani_Huse_Nahum_2018}%
  \BibitemOpen
  \bibfield  {author} {\bibinfo {author} {\bibfnamefont {V.}~\bibnamefont
  {Khemani}}, \bibinfo {author} {\bibfnamefont {D.~A.}\ \bibnamefont {Huse}},\
  and\ \bibinfo {author} {\bibfnamefont {A.}~\bibnamefont {Nahum}},\ }\bibfield
   {title} {\bibinfo {title} {Velocity-dependent lyapunov exponents in
  many-body quantum, semiclassical, and classical chaos},\ }\href
  {https://doi.org/10.1103/PhysRevB.98.144304} {\bibfield  {journal} {\bibinfo
  {journal} {Phys. Rev. B}\ }\textbf {\bibinfo {volume} {98}},\ \bibinfo
  {pages} {144304} (\bibinfo {year} {2018})}\BibitemShut {NoStop}%
\bibitem [{\citenamefont {Fishman}\ and\ \citenamefont
  {White}(2015)}]{Fishman_2015}%
  \BibitemOpen
  \bibfield  {author} {\bibinfo {author} {\bibfnamefont {M.~T.}\ \bibnamefont
  {Fishman}}\ and\ \bibinfo {author} {\bibfnamefont {S.~R.}\ \bibnamefont
  {White}},\ }\bibfield  {title} {\bibinfo {title} {Compression of correlation
  matrices and an efficient method for forming matrix product states of
  fermionic gaussian states},\ }\bibfield  {journal} {\bibinfo  {journal}
  {Physical Review B}\ }\textbf {\bibinfo {volume} {92}},\ \href
  {https://doi.org/10.1103/physrevb.92.075132} {10.1103/physrevb.92.075132}
  (\bibinfo {year} {2015})\BibitemShut {NoStop}%
\bibitem [{\citenamefont {Wu}\ \emph {et~al.}(2025)\citenamefont {Wu},
  \citenamefont {Kloss}, \citenamefont {Krinitsin}, \citenamefont {Fishman},
  \citenamefont {Pixley},\ and\ \citenamefont {Stoudenmire}}]{Wu_2025}%
  \BibitemOpen
  \bibfield  {author} {\bibinfo {author} {\bibfnamefont {A.-K.}\ \bibnamefont
  {Wu}}, \bibinfo {author} {\bibfnamefont {B.}~\bibnamefont {Kloss}}, \bibinfo
  {author} {\bibfnamefont {W.}~\bibnamefont {Krinitsin}}, \bibinfo {author}
  {\bibfnamefont {M.~T.}\ \bibnamefont {Fishman}}, \bibinfo {author}
  {\bibfnamefont {J.~H.}\ \bibnamefont {Pixley}},\ and\ \bibinfo {author}
  {\bibfnamefont {E.~M.}\ \bibnamefont {Stoudenmire}},\ }\bibfield  {title}
  {\bibinfo {title} {Disentangling interacting systems with fermionic gaussian
  circuits: Application to quantum impurity models},\ }\bibfield  {journal}
  {\bibinfo  {journal} {Physical Review B}\ }\textbf {\bibinfo {volume}
  {111}},\ \href {https://doi.org/10.1103/physrevb.111.035119}
  {10.1103/physrevb.111.035119} (\bibinfo {year} {2025})\BibitemShut {NoStop}%
\bibitem [{\citenamefont {Fukuda}\ \emph {et~al.}(2019)\citenamefont {Fukuda},
  \citenamefont {König},\ and\ \citenamefont {Nechita}}]{Fukuda_2019}%
  \BibitemOpen
  \bibfield  {author} {\bibinfo {author} {\bibfnamefont {M.}~\bibnamefont
  {Fukuda}}, \bibinfo {author} {\bibfnamefont {R.}~\bibnamefont {König}},\
  and\ \bibinfo {author} {\bibfnamefont {I.}~\bibnamefont {Nechita}},\
  }\bibfield  {title} {\bibinfo {title} {Rtni—a symbolic integrator for
  haar-random tensor networks},\ }\href
  {https://doi.org/10.1088/1751-8121/ab434b} {\bibfield  {journal} {\bibinfo
  {journal} {Journal of Physics A: Mathematical and Theoretical}\ }\textbf
  {\bibinfo {volume} {52}},\ \bibinfo {pages} {425303} (\bibinfo {year}
  {2019})}\BibitemShut {NoStop}%
\end{thebibliography}%

        \begin{appendices}

	\section{Weingarten calculation for Haar random unitaries}
	\label{app:Wg}
	In this appendix, we detail the derivation of the correlation spectrum for Haar random unitaries. We consider the elements of the matrix $\Delta$ as the averaged value over the Haar measure of the $d$-dimensional unitary group $U(d)$, first for the chosen operator to be the time-evolution operator:
	\begin{equation}
		\mathbb{E}_{U}(\Delta_{mn}^{}[\hat{U}]) = \int \mathrm{d}U \, \mathrm{Tr}\left[ U^{\dagger}\hat{c}^{\dagger}_{m} U \hat{c}^{}_{n} \right],
		\label{eq:weingarten_expect_U}
	\end{equation}
	and also for an evolving local operator $\hat{O} = U^{\dagger}\hat{n}_i^{}U$:	
	\begin{equation}
		\mathbb{E}_{U}(\Delta_{mn}^{}[\hat{n}^{}_{i}]) = \int \mathrm{d}U \, \mathrm{Tr}\left[ U^{\dagger}\hat{n}_i^{}U \hat{c}^{\dagger}_{m}  U^{\dagger}\hat{n}_i^{}U\hat{c}^{}_{n} \right],
		\label{eq:weingarten_expect_loc}
	\end{equation}
	The average over the
	unitary group is performed using Weingarten calculus, using its graphical representation. Following Ref.~\cite{Collins2010_Weingarten}, we can express the expected value in terms of
	diagrams $\mathcal{D}_r$ obtained after a removal procedure $r$ of the original diagram $\mathcal{D}$, chosen in the space of possible removals $r\in \mathrm{Rem}(\mathcal{D})$.
	The expectation value thus reads:
	\begin{equation}
		\mathbb{E}_{U}(\mathcal{D}) = \sum_{r=(\alpha,\beta)\in\mathrm{Rem}(\mathcal{D})}\mathcal{D}_r \mathrm{Wg}(d, \alpha\beta^{-1}), \
		\label{eq:removals}
	\end{equation}
	where  $\mathrm{Wg}(d, \sigma)$ is the Weingarten function, $d$ is the dimension of the matrix $U$, and $(\alpha,\beta)$ is a set of two permutations defining the corresponding
	removal $r$, which we explain further below. In the large-$d$ limit, the Weingarten function takes simple forms that we directly plug in our equations \cite{Collins2010_Weingarten,Fukuda_2019}.
	
	\begin{figure}
		\includegraphics[width=1.0\columnwidth]{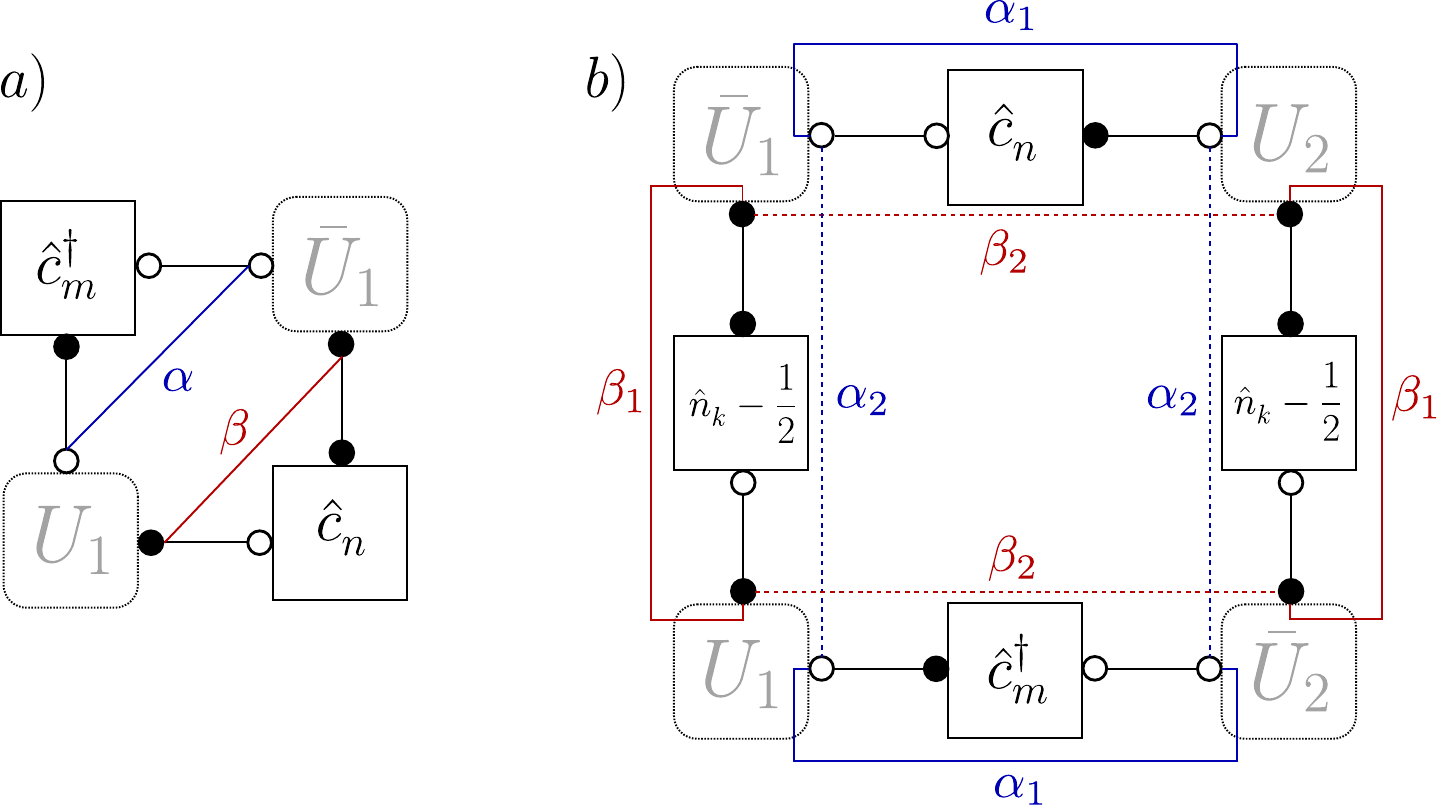}
		\caption{Weingarten diagram for the trace of operators $\mathrm{Tr}\left(\hat{O} \hat{c}^{\dagger}_{m} \hat{O}\hat{c}^{}_{n}\right)$, corresponding to the elements of the matrix $\Delta$ for Haar random unitaries. a) $\hat{O} = \hat{U}(t) $. b)  $\hat{O} = \hat{U}^\dagger(t)(\hat{n}_k-1/2) \hat{U}(t) $. We note $U^{\dagger} = \bar{U}^{t}$, where the transpose is performed by inverting the direct and dual spaces, denoted by empty and full circles respectively. Blue lines connect direct spaces of $U_{i}$ and $U_{\alpha(i)}$ and red lines connect the dual spaces of $U_{i}$ and $U_{\beta(i)}$. Solid and dashed lines correspond to the two different possible permutations for $\alpha$ and $\beta$.}
		\label{fig:weingarten_diagram}
	\end{figure}
	\emph{Removals}: In order to explain the way to perform the removal, we refer to Fig.~\ref{fig:weingarten_diagram}, which shows a diagrammatic representation of Eq.~\eqref{eq:weingarten_expect_U} and Eq.~\eqref{eq:weingarten_expect_loc} (panels a) and b) respectively). In these diagrams, white (resp. black) circles refer to the direct (dual) Hilbert space on which an operator (represented by boxes) act. We also
	performed the transpose of $\bar{U} = (U^{\dagger})^{t}$ by inverting black and white circles on the corresponding operators. In this case, $U_{1,2}$ and $\bar{U}_{1,2}$
	correspond to a same random realization of a random unitary matrix $U$, but the indices are useful to define a removal $r=(\alpha, \beta)$. $\alpha$ (resp. $\beta$) is a
	permutation connecting the white (black) circle of $U_{i}$ to the white (black) circle of $\bar{U}_{\alpha(i)}$ ($\bar{U}_{\beta(i)}$). Each combination of $\alpha$ and
	$\beta$ corresponds to a removal $r$: we connect the lines (vector spaces) according to the chosen permutations, and read what is left in $\mathcal{D}_r$. \\

	In the panel a) of Fig.~\ref{fig:weingarten_diagram}, there is only a single choice for $\alpha$ and $\beta$, leading to a single reduced diagram. The reduced diagram reads $ \mathcal{D}_r = \mathrm{Tr}(\hat{c}^{\dagger}_m) \mathrm{Tr}(\hat{c}^{}_n)=0$, since these operator only have off-diagonal elements. The elements of $\Delta$ are thus always $0$, such that the eigenvalues of $R$ are trivially all equal to $1/2$.

	In the panel b) of Fig.~\ref{fig:weingarten_diagram}, we have two choices for $\alpha$ (blue lines) and two for $\beta$ (red lines), which leads to four reduced diagrams. We summarize the
	different combinations in the following table:
	\begin{equation*}
		\begin{matrix}
			\alpha 	&&	\beta  &&  \mathcal{D}_r 				&& \mathrm{Wg}(d, \alpha\beta^{-1})\\
			\hline
			(12) 	&&	(1)(2) &&  \mathrm{Tr}(\hat{c}^\dagger_m) \mathrm{Tr}(\hat{c}^{}_n)	 \mathrm{Tr}(\hat{O})^{2}	&& -d^{-3} \\
			(12) 	&&	(12)   && \mathrm{Tr}(\hat{c}^\dagger_m) \mathrm{Tr}(\hat{c}^{}_n)\mathrm{Tr}(\hat{O}^{2})	&& d^{-2} \\
			(1)(2) 	&&	(1)(2) &&  \mathrm{Tr}(\hat{c}^\dagger_m\hat{c}^{}_n)\mathrm{Tr}(\hat{O})^{2}		&& d^{-2} \\
			(1)(2) 	&&	(12)   &&  \mathrm{Tr}(\hat{c}^\dagger_m\hat{c}^{}_n)\mathrm{Tr}(\hat{O}^{2})		&& -d^{-3}
		\end{matrix},
	\end{equation*}
	where $\hat{O} = \hat{n}_i-1/2$. Only $\mathrm{Tr}(\hat{c}^\dagger_m\hat{c}^{}_n)$ and $\mathrm{Tr}(\hat{O}^{2})$  are non zero, such that only the last element of the table above is not zero. This last term evaluates to $\mathrm{Tr}(\hat{c}^\dagger_m\hat{c}^{}_n)\mathrm{Tr}(\hat{O}^{2}) \mathrm{Wg}(d, \alpha\beta^{-1}) = -\delta_{mn}\frac{d}{2}\frac{d}{4}d^{-3}=-\delta_{mn}\frac{1}{8d}$. Hence the eigenvalues of $R$ are given by $\lambda_n = 1/2 \pm 1/(8d)$, where $d=2^L$ is the dimension of the corresponding Fock space. Local operators rotated by a random Haar unitary have therefore a non trivial correlation spectrum for all finite system sizes, but tend exponentially to the one of a fully correlated operator with system size.

	\section{Occupations of SNOs in the atomic limit}
	\label{app:Large_U}
	In the limit of large interaction strength $U\to\infty$, we assume that the impurity is disconnected from the bath and study the  OBDMO in the atomic limit. The
	Hamiltonian that we consider is:
	\begin{equation}
		H = U\left(\hat{n}_{1}-\frac{1}{2} \right)\left(\hat{n}_{2}-\frac{1}{2} \right) + V \left(\hat{c}^{\dagger}_{1}\hat{c}^{}_{2} + \hat{c}^{\dagger}_{2}\hat{c}^{}_{1}
		\right),
	\end{equation} 
	and the corresponding time-evolution operator takes the form of a $(4\times 4)$ matrix:
	\begin{equation}
		\begin{split}
			\left[\hat{U}(t)\right] &= \exp(itU
			\begin{bmatrix}
				0 & 0 & 0 & 0 \\
				0 & -\frac{1}{2} & \frac{V}{U} & 0 \\
				0 & \frac{V}{U} & -\frac{1}{2} & 0 \\
				0 & 0 & 0 & 0
			\end{bmatrix}
			) \\
			&= 
			e^{i\frac{\tilde{t}}{2}}
			\begin{bmatrix}
				e^{-i\frac{\tilde{t}}{2}} & 0 & 0 & 0 \\
				0 & \cos(\omega) & i\sin(\omega) & 0 \\
				0 & i\sin(\omega) & \cos(\omega) & 0 \\
				0 & 0 & 0 & e^{-i\frac{\tilde{t}}{2}}
			\end{bmatrix},
		\end{split}
	\end{equation}
	where $\omega=V/U$ and $\tilde{t}=Ut$. We can compute the elements of the matrix $\Delta$ defined as $[\Delta]_{mn} =
	\mathrm{Tr}[\hat{U}^{\dagger}(t)\hat{c}^{\dagger}_{m}\hat{U}(t)\hat{c}^{}_{n}]/\mathrm{Tr}[\mathds{1}]$, which reads:
	\begin{equation}
		\Delta = \frac{1}{2}
		\begin{bmatrix}
			\cos(\omega \tilde{t}) \cos(\frac{\tilde{t}}{2}) &  -\sin(\omega \tilde{t}) \sin(\frac{\tilde{t}}{2}) \\
			-\sin(\omega \tilde{t}) \sin(\frac{\tilde{t}}{2}) &  \cos(\omega \tilde{t}) \cos(\frac{\tilde{t}}{2})
		\end{bmatrix}.
	\end{equation}
	It is straightforward to find the eigenvalues of the matrix $\Delta^{\dagger}\Delta$, which are the following oscillating functions:
	\begin{equation}
		\mu_{\pm} = \frac{1}{4}\cos(\omega \tilde{t} \pm \frac{\tilde{t}}{2})^{2},
	\end{equation}
	from which we find the four eigenvalues of $R$ as $\lambda_n = 1/2\pm\sqrt{\mu_n} = 1/2 \pm \cos((V\pm U/2)t)/2$, as presented in Eq.~\eqref{eq:largeUocc}. Note that this equation
	allows to find the correct fast component of the oscillations, which is given by the dominating term $U$. However, the slow component is not correctly captured as it would
	require to introduce effects from the bath which are of the same order in $1/U$, since $V$ is not considered large here.
	\clearpage

	\end{appendices}

\end{document}